\documentclass[11pt]{article}
\RequirePackage{latexsym}
\RequirePackage{graphicx}
\RequirePackage{amssymb}
\RequirePackage{setspace}

\usepackage{color}
\usepackage{array}
\usepackage{url}

\begin{document}

\definecolor{grey}{rgb}{0.5,0.5,0.5}
\definecolor{gold}{rgb}{0,0,0}
\definecolor{blue}{rgb}{0,0,0}
\definecolor{red}{rgb}{0,0,0}
     \begin{figure}
     \begin{center}
     \includegraphics[scale=0.8]{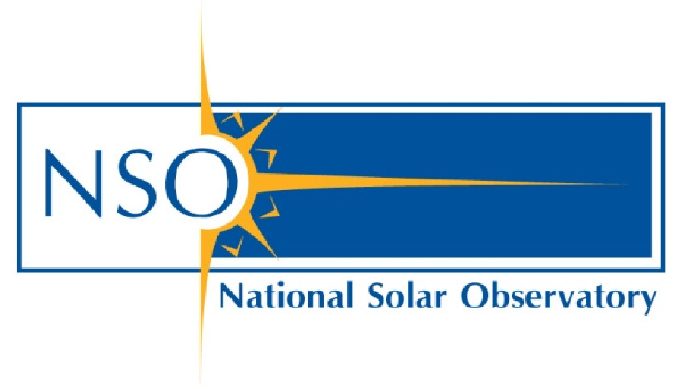}
     \end{center}
     \end{figure}
\title{{\bf Solar P-angle Alignment in GONG Dopplergrams}
\\$\,$}
\author{Anna L.~H.~Hughes, Irene Gonz\'alez-Hern\'andez\footnote{Deceased}, Sean G. McManus\footnote{Present Addesss: NOIRLab, Tucson, AZ} , Kiran Jain\\
and Sushanta C. Tripathy
\\ NISP, National Solar Observatory, Boulder, CO
  \\$\,$  \\$\,$}
\maketitle

\noindent\rule{\linewidth}{0.5mm}

\begin{center}
Technical Report No. {\bf NSO/NISP-2013-004, Version 2}
\\$\,$
\end{center}

\begin{abstract}
\addcontentsline{toc}{subsection}{\textcolor{blue}{\bf Abstract}}

In helioseismic studies, an observational parameter of primary concern is the P-angle, the angle along which lies the solar axis of rotation for a given image.  For the six observing sites employed by The Global Oscillation Network Group (GONG), this angle acts additionally as a marker of relative image orientation, allowing concurrent images to be precisely aligned and merged to provide the highest possible quality data. In this report, we present and investigate two methods of determining the P-angle via the rotational signature embedded in solar Dopplergram images by examining the large-scale structure of the observed velocity field.  As with other studies, we find that the Dopplergram produces a time-varying 'P-angle' signature according to the presentation of various physical phenomena across the solar surface, but with the potential for sub-degree identification of the axis of rotation.  However, close agreement between separate P-angle-finding techniques also reveals current limitations to P-angle determination that are imposed by the calibration state of the GONG-site Dopplergrams, leaving these P-angle-finding methods for GONG 
with errors on the scale of less than a degree between two site. 

\end{abstract}

\pagebreak
\tableofcontents

\section[\textcolor{blue}{Introduction}]{\textcolor{blue}{Introduction}}
\label{Introduction}

\begin{figure}
     \begin{center}
     \includegraphics[scale=0.74]{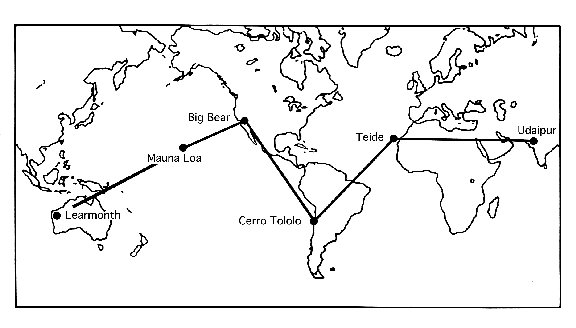}
     \caption[Sites of GONG]{Six sites of the GONG network: Learmonth (LE), Mauna Loa (ML), Big Bear (BB), Cerro Tololo (CT), Teide (TD) and Udaipur (UD).}
     \label{GONG_sites}
     \end{center}
     \end{figure}

The Global Oscillation Network Group (GONG) is a six-site network, with instruments deployed around the world dedicated to providing 24-hour coverage of the Sun (see Figure~\ref{GONG_sites}).  Used primarily for helioseismology, it operates under tight tolerances when it comes to the merging of concurrent site data and to the orientation of data images.  Of primary importance is the solar P-angle, which defines the angle of solar-rotational true-north for each observation.  Due to the extreme sensitivity of local helioseismology applications to P-angle errors, GONG processing strives for an alignment accuracy of better than 0.02 degrees.

The current GONG alignment proceedure relies on each site taking regular calibration observations, which result in site outages during the periods of calibration.  Further, these calibrations act as markers, but, as the observed P-angle varies for each site by about 0.25 degrees daily with further seasonal and maintenance variations, they cannot define the precise alignment during all times of GONG operation.  To make up the difference, an IRAF (Image Reduction and Analysis Facility)-based cross-correlation algorithm called xoffset is used to closely align the fine-scale structure observed within concurrent Dopplergrams \cite{TonerHarvey1998}.  This alignment is then referenced against the site with the most recent and complete calibration observations to anchor the network alignment as a whole.  The alignment processes are coordinated by the GONG reduction stage COPIPE \cite{COPIPEwebdoc, IRAFcamoffset}.

Unfortunately, while the GONG alignment methods have worked well in the past and for past helioseismology applications, these alignment methods are quite implementation dependent as well as being vulnerable to large-scale network outages and to instances when strings of bad calibration sets occur.  Therefore, we would like to explore other methods that may offer useful alternatives to image-alignment and P-angle identification.

The Michelson Doppler Imager (MDI) flew aboard the {\em Solar and Heliospheric Observatory} ({\em SOHO}) spacecraft for about fifteen years, providing a steady stream of Doppler and magnetic-field images of the sun.  During a period of mechanical failure, the SOHO operators performed an investigation into using a simple Dopplergram-based algorithm to inspect the spacecraft roll-angle via the solar P-angle \cite{MDIDopHalf}.  Review of their documentation suggests that, when results were averaged over an hour's worth of data, they could identify the P-angle to within about 0.5 degrees.  Therefore, in this report, we begin an investigation into possible improvements to the SOHO/MDI algorithm as well as other Dopplergram-based methods to discover whether the rotational signature encoded within these data sets can provide the GONG network with the alignment information it needs.

The layout of the rest of this report is as follows:
\begin{list}{}{}
     \item \S\ref{Algorithms}: Details of the various algorithms used in this study, both those currently in use in COPIPE and those under investigation as future potential tools.
     \item \S\ref{Data}: Outline of various pieces of information necessary for working with GONG Dopplergrams.
     \item \S\ref{Results}: Overview of results that were derived during the development and testing of the Dopplergram-P-angle algorithms.
     \item \S\ref{Conclusions}: List of conclusions we have achieved with the present study.
     \item \S\ref{Future}: Outline of a number of avenues the remain open for future investigation and/or refinement of the Dopplergram-P-angle algorithms.
\end{list}

\section[\textcolor{blue}{Algorithms Tested and Compared Against}]{\textcolor{blue}{Algorithms Tested and Compared Against}}
\label{Algorithms}

There are four different P-angle--associated algorithms/values discussed in this report.  The first two are the xoffset routine, and the OFFSET value reported in the fits header of the Level-2 P-angle corrected GONG-site velocity images.  These are currently in use within the GONG COPIPE processing suite and are used in this report primarily as the fiducials against which we compare the results of our experimental algorithms.  The other two are the DopHalf algorithm and the RingPhase algorithm, which are the primary subjects of investigation in this report.  Both of these algorithms take in a single, raw, site-specific Dopplergram image and examine the large-scale structure of the velocity field in order to estimate the orientation of the solar axis of rotation.

     \subsection[\textcolor{blue}{xoffset}]{\textcolor{blue}{xoffset}}
     \label{Algorithms_xoffset}

The xoffset routine is written for the GONG Reduction and Analysis Software Package (GRASP) within IRAF and has the task of computing alignment angles between concurrent observation-images from different sites \cite{IRAFxoffset}.  This image-registration algorithm takes in Dopplergram images, but the underlying algorithm is compatible with any solar-image--data type.  Dopplergrams, however, have a high degree of fine structure, allowing for tighter alignment precision.

\begin{figure}
     \begin{center}
     \includegraphics[scale=0.18]{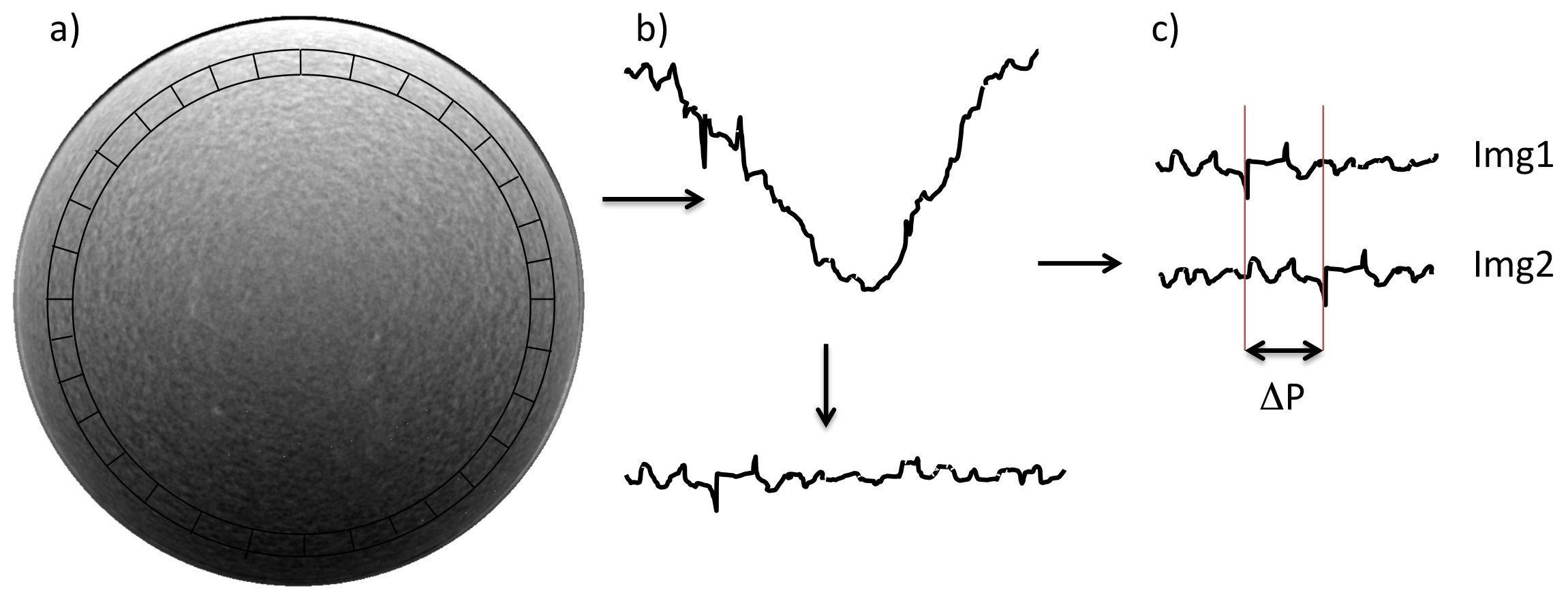}
     \caption[The xoffset view of Dopplergrams]{Cartoon of the xoffset treatment of solar Dopplergrams.}
     \label{FIG_Algoxoffset}
     \end{center}
     \end{figure}

The basic processes of the xoffset algorithm are diagramed in Figure~\ref{FIG_Algoxoffset}
and proceed as follows:
\begin{list}{* }{}
     \item Several equal-area annuli are defined near the solar limb.
     \item For each annulus, the Dopplergram-velocities are binned into 1024 angle bins, creating a velocity curve as a function of angle for a given radius.
     \item The velocity curve is filtered, removing the long-wavelength components and leaving only the fine-structure of the velocity field for high-precision alignment.
     \item The filtered curve is compared to the corresponding curve taken from a concurrent image using 1D-Fast-Fourier-Transform (FFT) cross-correlations.  The cross-correlation returns the offset of the two curves, which is equivalent to the rotational offset or P-angle difference between the two images.
     \item The P-angle differences calculated for each of the sampled annuli are averaged to return the final difference-angle result for the two images.
\end{list}
Note that in order for this algorithm to function, the solar-center coordinates and limb boundary must be defined quite accurately for each image.  Also, xoffset is designed to handle somewhat elliptical solar images.  In terms of results, xoffset tends to report merit values for concurrent GONG-image alignments in the range of 0.97 and uncertainties of less than $0.01^\circ$.  See Toner and Harvey \cite{TonerHarvey1998} for a more formal description and error analysis.

     \subsection[\textcolor{blue}{OFFSET and Drift-scans calibration}]{\textcolor{blue}{OFFSET and Drift-scans calibration}}
     \label{Algorithms_OFFSET}

\begin{figure}
     \begin{center}
     \includegraphics[scale=0.08]{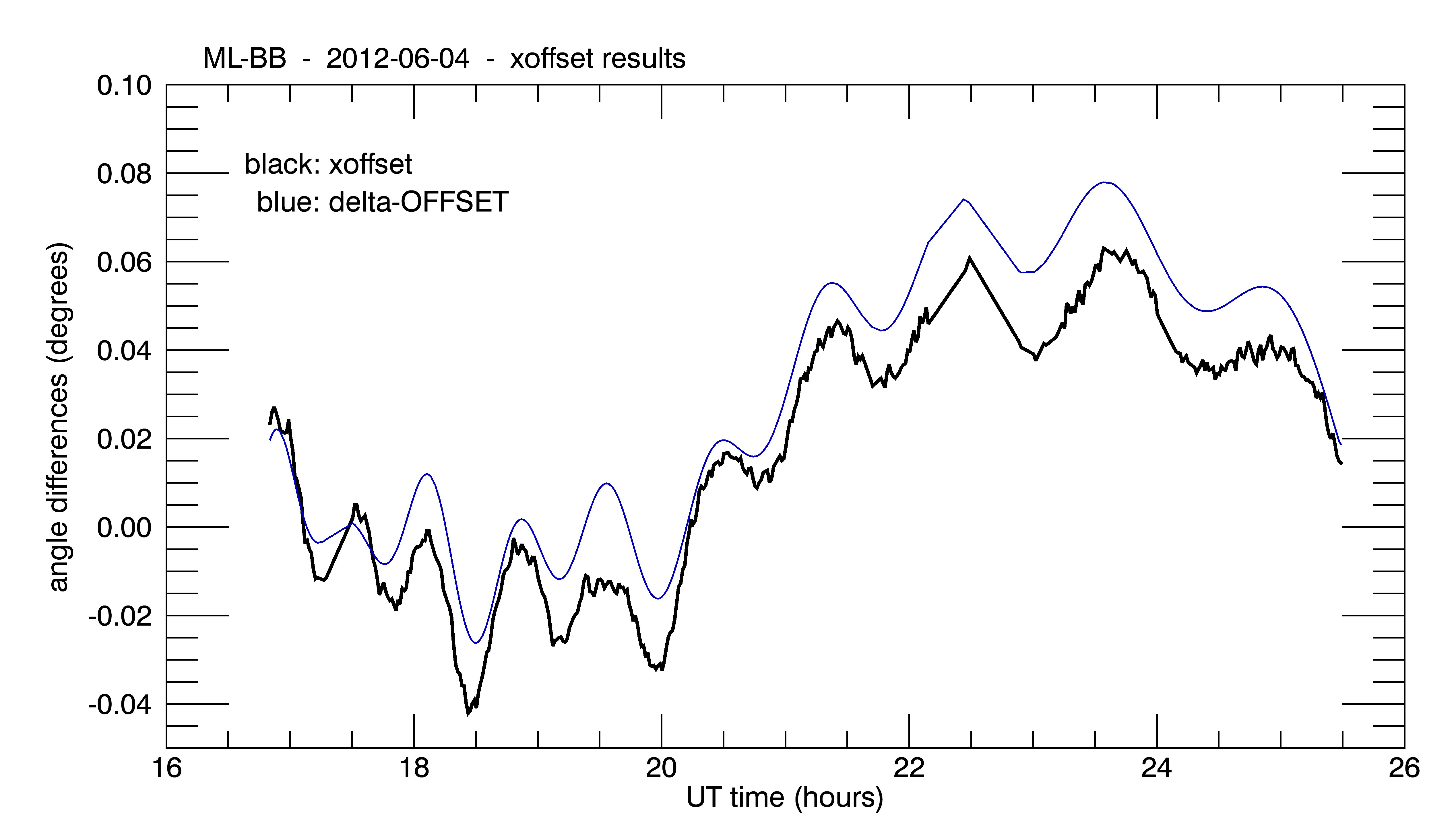}
     \caption[OFFSET vs. xoffset]{Comparison of the xoffset results obtained between a set of Big Bear and Mauna Loa site images to the difference of the OFFSET angles reported in the two sites' fits headers.}
     \label{FIG_AlgoOFFSETvxoffset}
     \end{center}
     \end{figure}

OFFSET is a keyword reported in the Level-2 GONG-site images, which provides the COPIPE-calculated best estimate of the P-angle for a given observation.  A lot of different elements go into calculating OFFSET.  For instance, the results of xoffset for individual image pairs are used to reference the P-angles of severel GONG sites against one site who's P-angle values throughout a day are the most up-to-date and currently well-known.  This relationship can be seen in Figure~\ref{FIG_AlgoOFFSETvxoffset}
comparing a time-series of xoffset results to the reported OFFSET-value differences between two sites.

\begin{figure}
     \begin{center}
     \includegraphics[scale=0.16]{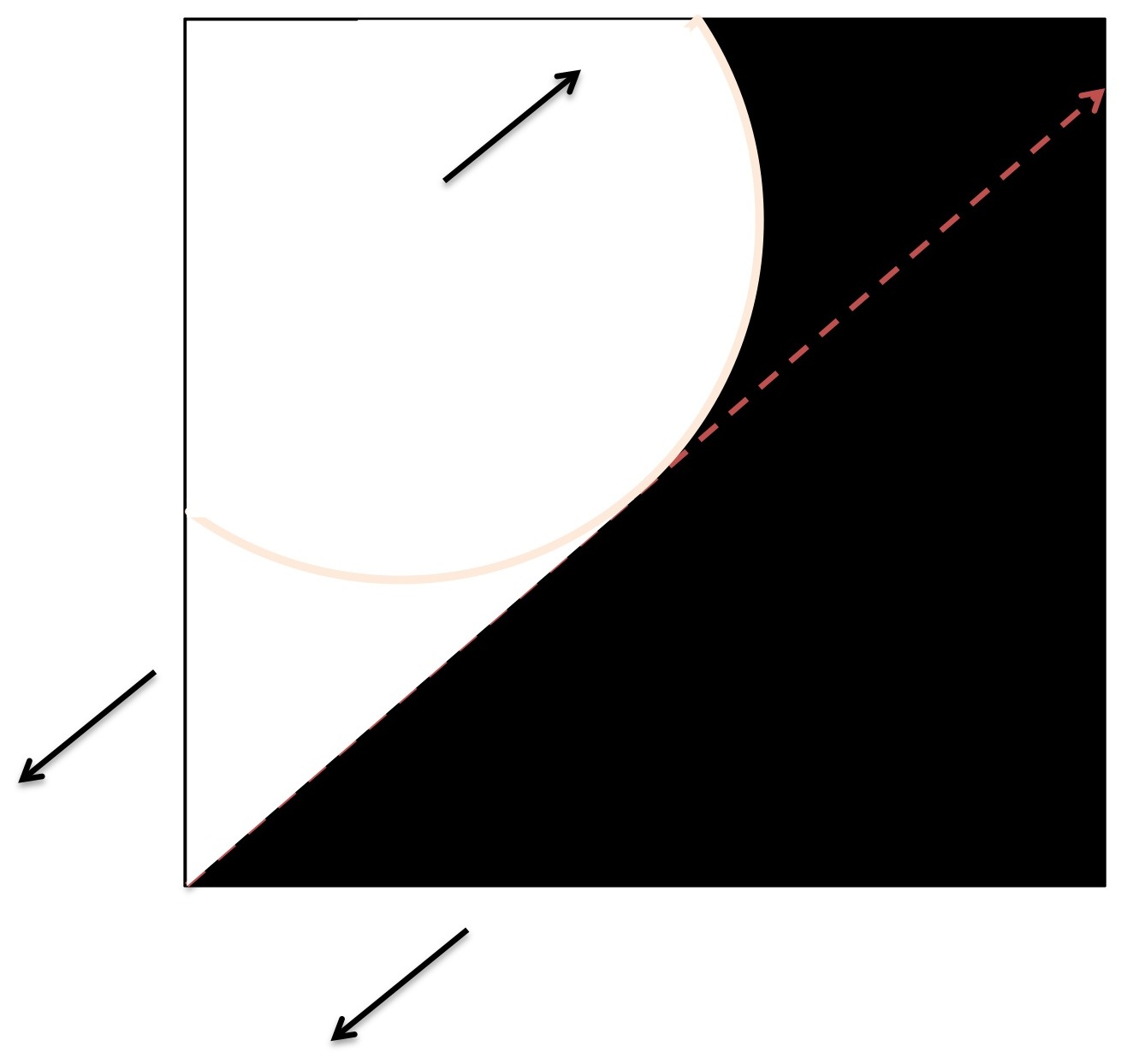}
     \caption[The Drifts-scans observations]{Cartoon of the method for acquiring Drift-scans calibration observations.}
     \label{FIG_AlgoDriftScans}
     \end{center}
     \end{figure}

At the heart of the OFFSET values, though, are the alignment-calibration observations known as Drift scans \cite{Driftscanwebdoc}.  Drift scans are taken, as depicted in Figure~\ref{FIG_AlgoDriftScans},
by fixing the camera and allowing the Sun to drift across/out of the image at different times of day.  This provides the terrestrial East-West vector to high precision, from which the P-angle can then be deduced.  Unfortunately, while the P-angle varies by up to $0.25^\circ$ throughout a day, these scans are time-consuming to acquire, resulting in a GONG-site outage while they occur.  Therefore, full-day scans for each site are taken only annually on a rotating basis with the other sites, while noon-drifts to calibrate the bias value of the full-day curve are taken every six to twelve days.  See Toner \cite{Toner2001} for a full description of the Drift scans procedure.

     \subsection[\textcolor{blue}{DopHalf}]{\textcolor{blue}{DopHalf}}
     \label{Algorithms_DopHalf}

At its core, the DopHalf algorithm is a direct replica of the algorithm that was used as a calibration anchor to estimate the roll-angle of the {\em SOHO} spacecraft via the solar-axis of rotation (the P-angle) observed in the MDI Dopplergrams \cite{MDIDopHalf}.  This method is viable using solar Dopplergrams only, and for GONG using only site images where the solar-rotational signal has not been removed.  The algorithm estimates the P-angle by examining only the largest-scale gradients in the velocity image, by comparing the mean velocities computed for the northern, southern, eastern, and western solar-image hemispheres.

\begin{figure}
     \begin{center}
     \includegraphics[scale=0.22]{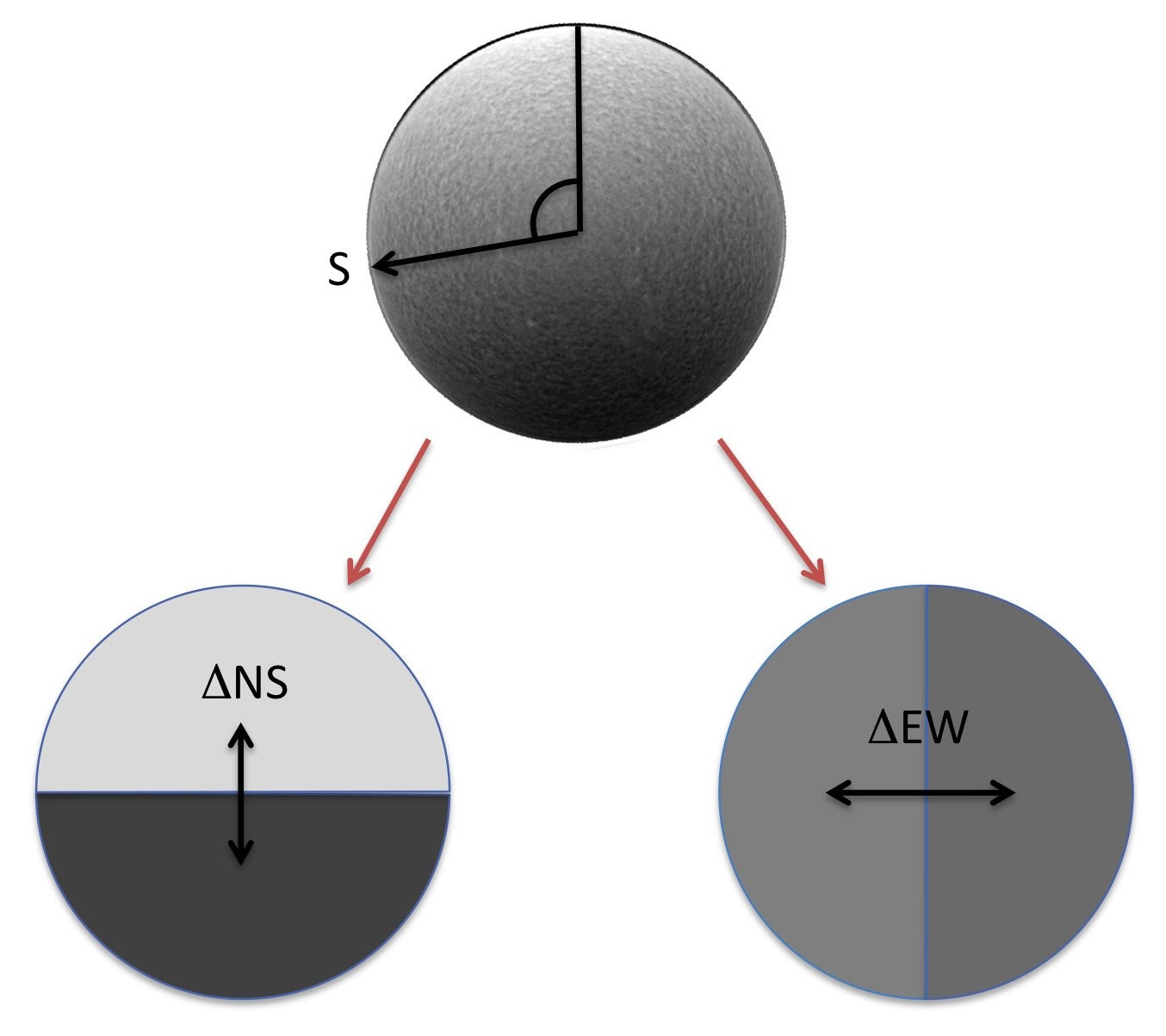}
     \caption[The DopHalf view of Dopplergrams]{Cartoon of the DopHalf treatment of solar Dopplergrams.}
     \label{FIG_AlgoDopHalf}
     \end{center}
     \end{figure}

For a solar image with a rotational axis oriented perfectly along the y-axis (0 or 180-degrees P-angle), only the eastern- and western-image hemispheres should report a difference in their mean velocities.  Likewise, if the rotation-axis is oriented perfectly along the x-axis (+/- 90-degrees P-angle), then only the northern and southern hemispheres should report a difference in their means.  Therefore, the equation used to compute the P-angle for the DopHalf algorithm is given by:
\begin{equation}
   \theta_P
=
   \arctan \left( \frac{\Delta NS}{\Delta EW} \right)
   \,,
\label{EQ_AlgoDopHalfP}
\end{equation}
where, as demonstrated in Figure~\ref{FIG_AlgoDopHalf},
$\Delta NS$ is the mean velocity of the northern-image hemisphere minus the mean velocity of the southern hemisphere, and $\Delta EW$ is the mean velocity of the western-image hemisphere minus the mean velocity of the eastern hemisphere.  {\em However}, because the presentation of individual GONG site images is flipped east-to-west relative to a traditional sky image, the above equation actually solves for the Solar {\em south} pole in these data.  We must therefore correct by $180^\circ$ to account for this fact when solving for $\theta_P$ --- measured {\em counter-clockwise} from the +y-axis.

Next, Equation~(\ref{EQ_AlgoDopHalfP}) holds true for a perfectly symmetrically-rotating sun, but in reality there are a number of various solar physical phenomena that contribute noise to this type of calculation.  As documented for the {\em SOHO} roll-angle estimate \cite{MDIDopHalf}, solar P-mode oscillations produce minute-to-minute variations on the order of 0.2 degrees, and longer-time-scale variations arise due to effects like the rotation of supergranules and of solar active regions.

For the purposes of this study where we examine ways to use, or possibly improve, this P-angle estimate, there are a few other points to note.  First is that we have not used this technique on only the raw Dopplergram images.  In several tests we apply an FFT filter to the image before computing hemispheric means.  See \S\ref{Results_DopHalf_Filtered} for details.

Second, the sampling that we use to calculate hemispheric means always adheres to the boundaries of the solar disk defined in the fits header, even in the cases of extreme image filtering.  In some cases, a buffer zone near the solar limb is excluded.  However, the hemispheric-mean calculation always uses the equation,
\begin{equation}
   V_\mathrm{mean}
=
   \frac{ \left( V_\mathrm{hem} * N_\mathrm{hem} 
                  + V_\mathrm{line} * N_\mathrm{line} * f_\mathrm{h}
             \right) }
           { \left( N_\mathrm{hem} + N_\mathrm{line} * f_\mathrm{h} \right) }
   \,,
\label{EQ_AlgoDopHalfVmean}
\end{equation}
where $V_\mathrm{mean}$ is the full hemispheric-mean calculated for a given hemisphere, $V_\mathrm{hem}$ is the mean value of only those {\em whole} pixels of a given image-hemisphere whose center positions fall within the limb radius or limb-buffer-reduced radius from the solar-image center, $N_\mathrm{hem}$ is the number of those whole pixels so selected, $V_\mathrm{line}$ is the mean value of those pixels in the pixel row or column containing the solar-center coordinates, $N_\mathrm{line}$ is the number of those pixels, and $f_\mathrm{h}$ is the width-fraction of that row or column that falls within the given hemisphere.

Finally, the solar-image disk as defined in the input fits header is slightly elliptical (see \S\ref{Data_PreFits}).  For this report, we have conducted DopHalf tests both adhering to this ellipticity in the radius and using the semi-minor-axis value to sample a circular disk.

     \subsection[\textcolor{blue}{RingPhase}]{\textcolor{blue}{RingPhase}}
     \label{Algorithms_RingPhase}

\begin{figure}
     \begin{center}
     \includegraphics[scale=0.18]{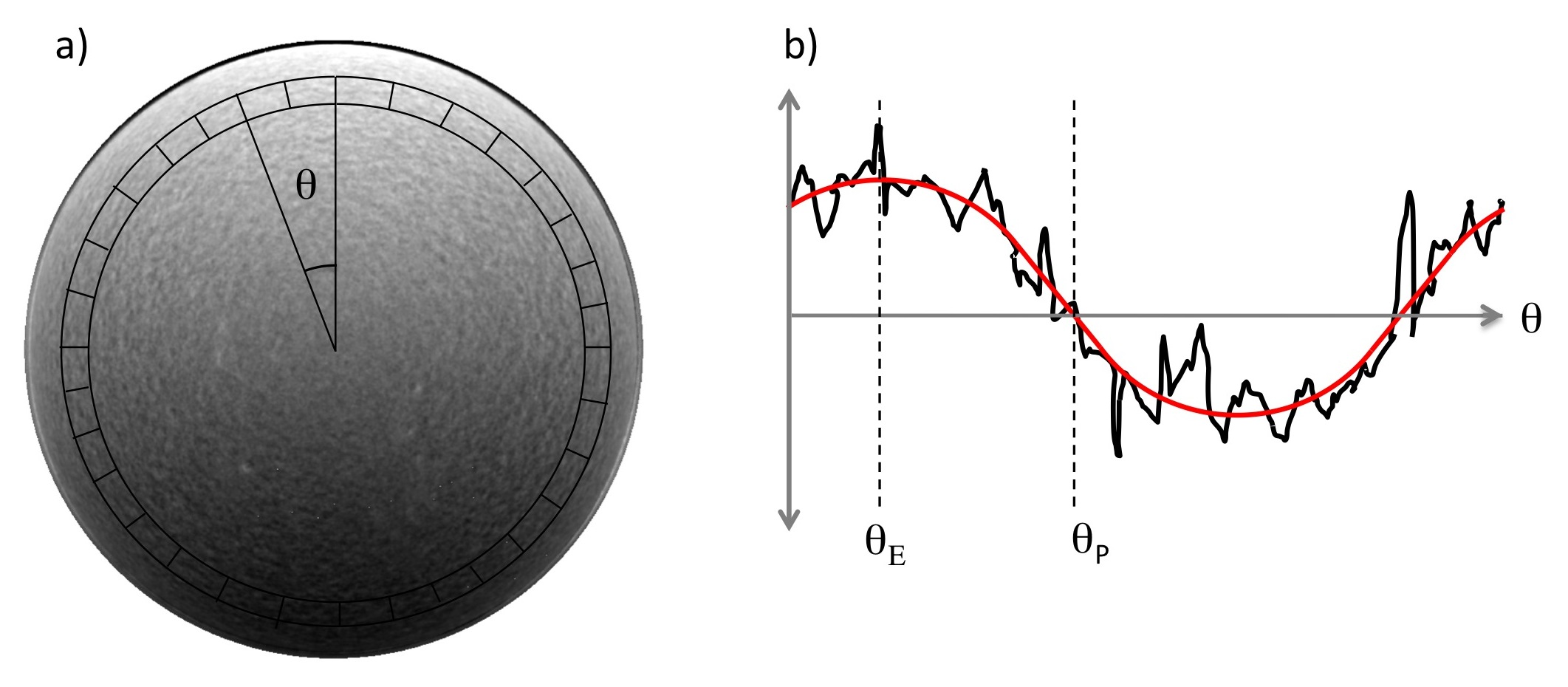}
     \caption[The RingPhase view of Dopplergrams]{Cartoon of the RingPhase treatment of solar Dopplergrams.}
     \label{FIG_AlgoRingPhase}
     \end{center}
     \end{figure}

The RingPhase algorithm for computing the solar P-angle uses only raw Dopplergram data and depends on 1D FFT phase angles to locate the solar equator-verus-poles orientation.  Like the xoffset routine, and as shown in Figure~\ref{FIG_AlgoRingPhase}, 
this algorithm examines annuli of velocity data binned into angle-bins with respect to the solar-image center.  However, where the xoffset routine discards the long-wavelength (solar-rotation-driven) information in these curves, RingPhase, like DopHalf, {\em depends} on the long-wavelength rotational signal.

The velocity curve is generated by computing the mean-value of the velocity in each annulus-angle bin (or, if the sampling is too fine, the bilinearly-interpolated value of the bin center position).  Once a velocity-versus-angle curve is generated, the code computes the 1D FFT, and from there computes the P-angle as:
\begin{equation}
   \theta_P
=
   \phi_1 - \frac{1}{2}\Delta\theta + 90^\circ
   \,,
\label{EQ_RingPhase}
\end{equation}
where $\phi_1$ is the phase-angle of the first (single-period in $360^\circ$) FFT component, and $\Delta\theta = 360^\circ / n_\theta$, the width of a single angle bin.  The $\frac{1}{2}\Delta\theta$ terms corrects the position of the phase angle relative to the starting-bin center position.  The equator on the sky-plane--western limb is positioned at  $-\phi_1 + \frac{1}{2}\Delta\theta$ in these images, measured counter-clockwise from the +y-axis.  However, Equation~\ref{EQ_RingPhase} provides $\theta_P$ {\em clockwise} from the +y-axis for ease of comparison with the header OFFSET values with a $+ 90^\circ$ to move from the western limb to the north pole in the orientation scheme of the east-west flipped GONG site images (see \S\ref{Data_Format}).

This algorithm is very much designed for use on perfectly circular, happily oriented solar Dopplergrams.  When used on elliptical images, it may require a correction factor, which is discussed in \S\ref{Data_Ellipses}.  Also, depending on the time-of-year that the data were taken, the solar image will usually not, in fact, present a lovely, even $90^\circ$ from equator to pole to equator to pole.  This is discussed further in \S\ref{Data_B0}.

Finally, in standard operation, the RingPhase algorithm reports the mean value of $\theta_P$ calculated for a number of equal-area annuli ranging, in this report, across the the full-radial-range of the solar disk.  These annuli are not exclusive, but instead overlap each other such that the starting radius of a given annulus is defined at the half-way point of the annulus interior to it.  The reported mean, therefore, somewhat underweights the contributions from the innermost and outermost annuli.

\section[\textcolor{blue}{Working with GONG Site Data}]{\textcolor{blue}{Working with GONG Site Data}}
\label{Data}

As with all real observations, there are a number of conventions and pitfalls associated with the GONG site data that one needs to be aware of in order to use the data effectively and accurately.  In the subsections below, we discuss the basic template for GONG site observations, the image-data fits and scalings pre-determined by the GONG early-stage processing software, and a couple of important issues to keep in mind related to the 3D orientation of the sun with respect to the telescopes.

In terms of the setup of the network itself, GONG has six science sites observing the sun from around the world.  These sites are at Mauna Loa in Hawaii (ML), Big Bear in California (BB), Cerro Tololo in Chile (CT), del Teide in the Cannary Islands (TD), Udaipur in India (UD), and Learmoth in Australia (LE).  There is also an engineering site that is sometimes used in Tucson, Arizona (TC), but its data are not included in the general repository of merged data.

     \subsection[\textcolor{blue}{Format and Orientation of Observations}]{\textcolor{blue}{Format and Orientation of Observations}}
     \label{Data_Format}

\begin{figure}[t]
     \begin{center}
     \includegraphics[scale=0.22]{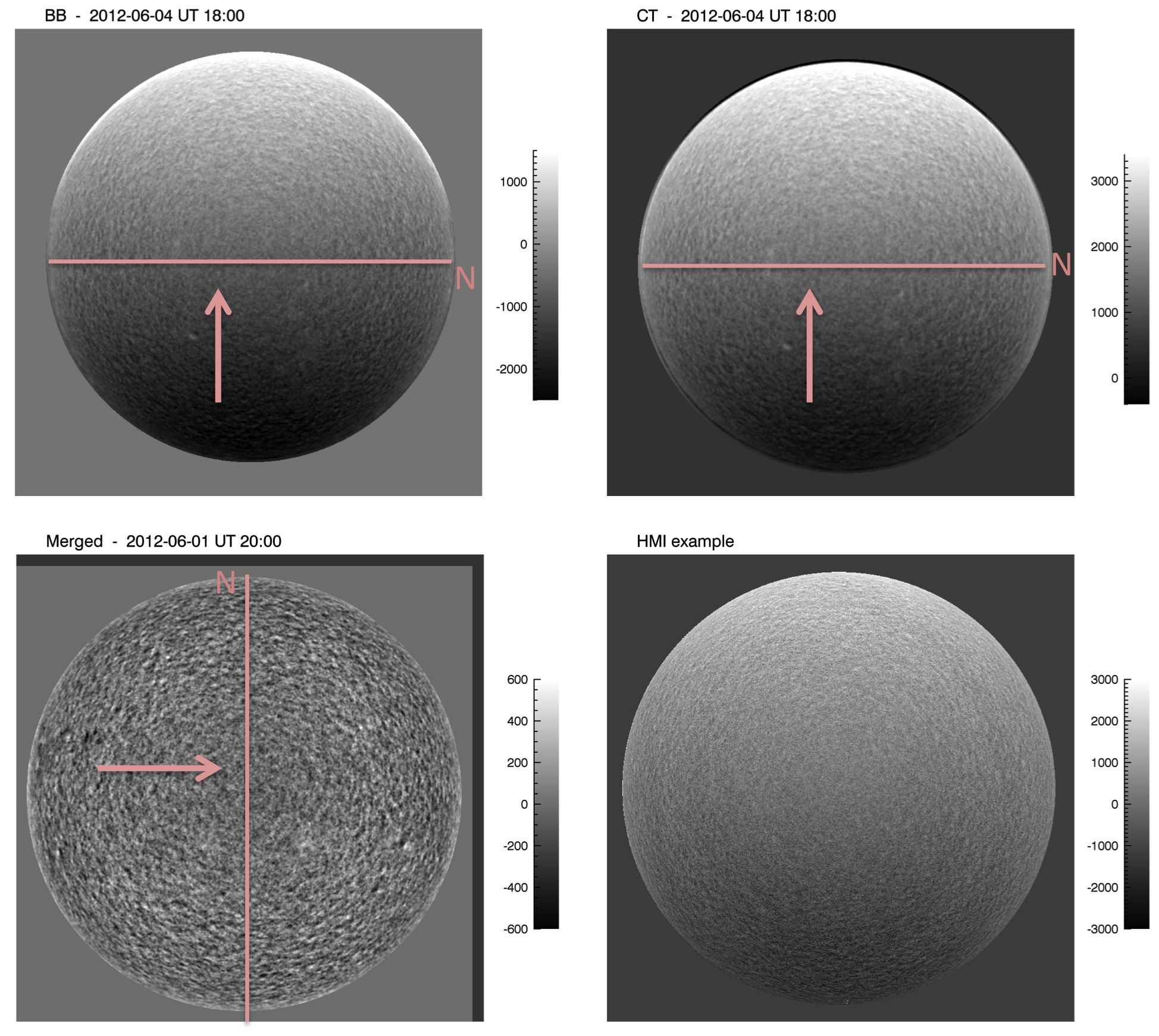}
     \caption[Examples of solar Dopplergrams]{Examples of solar Dopplergrams, with overlays for the GONG examples indicating the orientation of the Solar axis of rotation.  Arrows indicate the direction features in a time-series will rotate.  The top two panels present Dopplergrams from individual GONG sites.  The bottom-left panel presents a post image-merge GONG Dopplergram.  The bottom-right panel presents an example of a size-scaled HMI Dopplergram.  Only the two GONG-site examples are concurrent.}
     \label{FIG_DataExamples}
     \end{center}
     \end{figure}

The images displayed in Figure~\ref{FIG_DataExamples}
present examples of Dopplergrams as provided individually from two different GONG sites (in this case, Big Bear and Cerro Telolo, top panels), from a GONG multi-site--merged image (bottom-left panel), and, for comparison, from a scaled image taken from HMI data (bottom-right panel).  From these images one can see that the fiducial layout of the {\em site} images is for the solar-north pole to be oriented along the +x-axis, with the solar disk imaged at a little under 400 pixels in radius.  We also see that the individual site images are not scaled to the same velocity baseline due to local-observational and other effects, and also that the site images are flipped in the east-to-west so that in a time series features appear to have a left-handed rotation relative to the Solar North pole.

The merged data, on the other hand, have both had the velocity baselines normalized and the rotational profile fit and removed, the later of which processes makes the merged data unsuitable for use in our Dopplergram-based P-angle-finding algorithms.  They have also been re-oriented so that Solar North is at the top and flipped so that the east limb is naturally to the left and features in a time series move from left to right.  (Remember that in the plane of the sky the eastward limb of the Sun appears to the left if North is at the top.)  The full process for merging the Dopplergram images is discussed in a paper by Toner, et al. \cite{Toneretal2003}.

     \subsection[\textcolor{blue}{Important Keywords and Fits to the GONG-site Dopplergrams}]{\textcolor{blue}{Important Keywords and Fits to the GONG-site Dopplergrams}}
     \label{Data_PreFits}

In the case of the GONG individual-site Dopplergrams, there are two categories of data-fit information provided in the fits header that are important for the proper analysis of the data.  The first has to do with the translation of the values in the fits file into physically meaningful data.  The second provides the information on the positions of the solar center and limb in each image.

For the data translation, the important keywords are VELSCALE, VEL\_BIAS, and, optionally, VCOR1.  VELSCALE and VEL\_BIAS are computed for each site day "by fitting the mean velocity signal 
to the disk center velocity as computed from an ephemeris (JPL ephemerides are used)" \cite{CliffImmerge}.  See \cite{VELFIThelp} for further details.  In order to work with site Dopplergrams, the data should be read in from the fits file, and then translated according to:
\begin{equation}
   V_\mathrm{use}
=
   V_\mathrm{read-in} * \mathrm{VELSCALE} \,-\, \mathrm{VEL\_BIAS}
   - \mathrm{VCOR1}
\, ,
\end{equation}
(where the VCOR1 term is included only if it is necessary to correct for the difference between the observer frame and the solar-disk rest frame.)
While the algorithms presented in \S\ref{Algorithms_DopHalf} and \ref{Algorithms_RingPhase} are fundamentally scaling independent, if the given VELSCALE is $<0$, failure to employ the above data translation will lead to a computed P-angle $180^\circ$ off from where it should be.

\begin{figure}
     \begin{center}
     \includegraphics[scale=0.22]{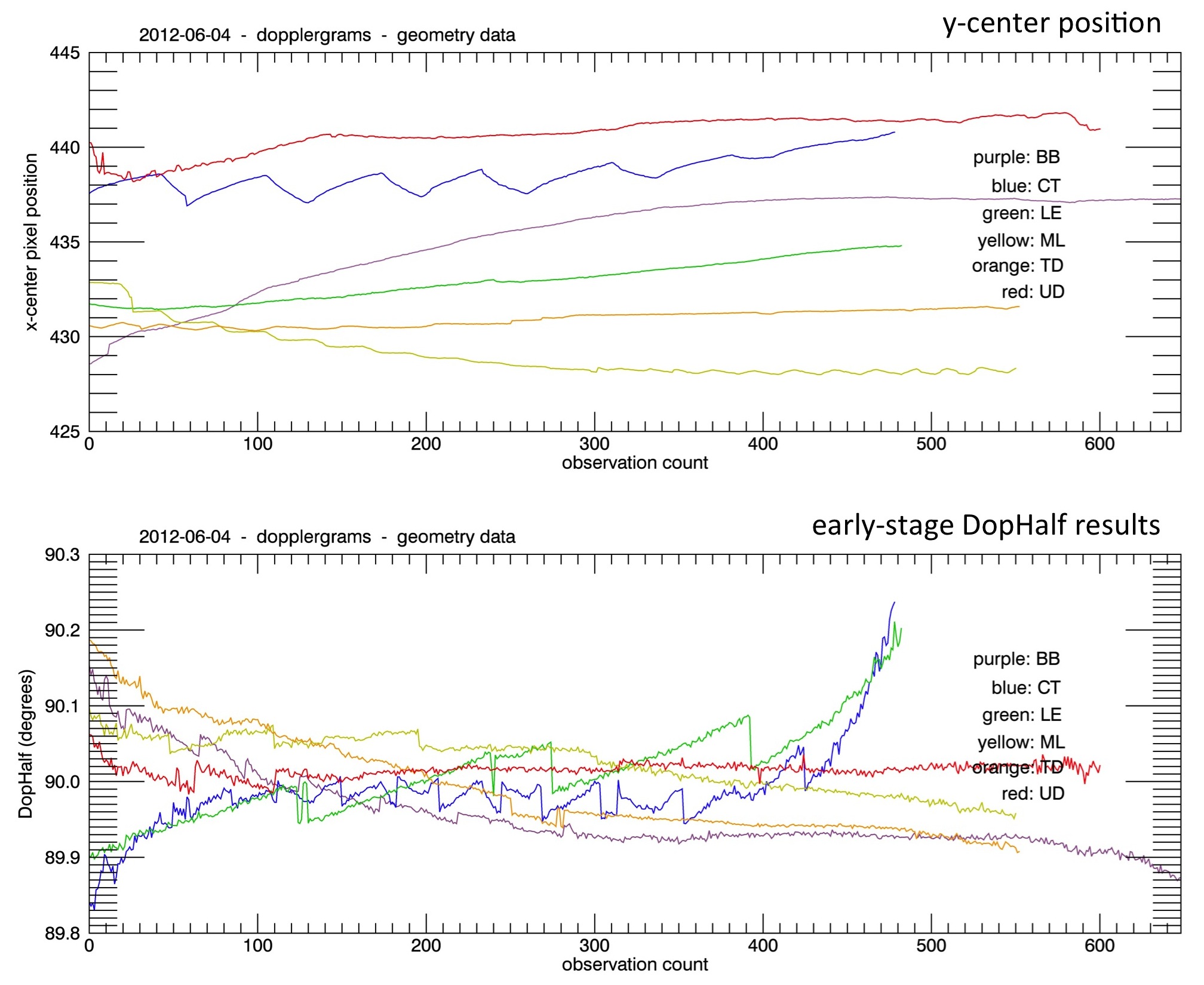}
     \caption[Solar-center pointing variations and effects]{The top panel demonstrates the pointing stability of GONG observations taken at different sites by displaying the computed solar-image y-center position throughout one day.  The bottom panel demonstrates the ways in which pixel-level pointing may manifest in buggy algorithms by providing the concurrent results from an early, test version of the DopHalf code.  In this case, the code was improperly handling the incorporation of the central row/column pixels into the hemisphere mean values.}
     \label{FIG_DataycenterVDopHalf}
     \end{center}
     \end{figure}

Next: the fit definitions of the solar-center coordinates and the fit defining the solar limb in each image.  The center coordinates for each solar image are given by the fits-header keywords FNDLMBXC and FNDLMBYC.  Note that any algorithm for determining the solar P-angle is going to be innately sensitive to any errors in these defined coordinates.  As can be seen in the top-panel of Figure~\ref{FIG_DataycenterVDopHalf},
the solar-center position in the GONG-site images usually, but not always, varies fairly smoothly.  However, as can be seen in the early DopHalf results presented in the bottom panel, instances of variable pointing {\em can} be used to detect pixel-fraction-handling errors in an algorithm, when movement across pixel cells produces jumps in computed results.

Finally, due to atmospheric refraction, the collected GONG site observations are ever-so-slightly elliptical, depending on the observing geometries at the time and place the data were gathered.  In order to define the solar limb in the image, we therefore need {\em three} other fits-header keyword parameters:  The best-fit for the semi-major axis is given by C\_MA, and for the semi-minor axis by C\_MI.  The orientation of the solar-limb ellipse is given by FNDLMBAN and is the counter-clockwise angle from the +y-axis to the semi-major axis of the ellipse \cite{IRAFfndlmb}.

\begin{figure}
     \begin{center}
     \includegraphics[scale=0.28]{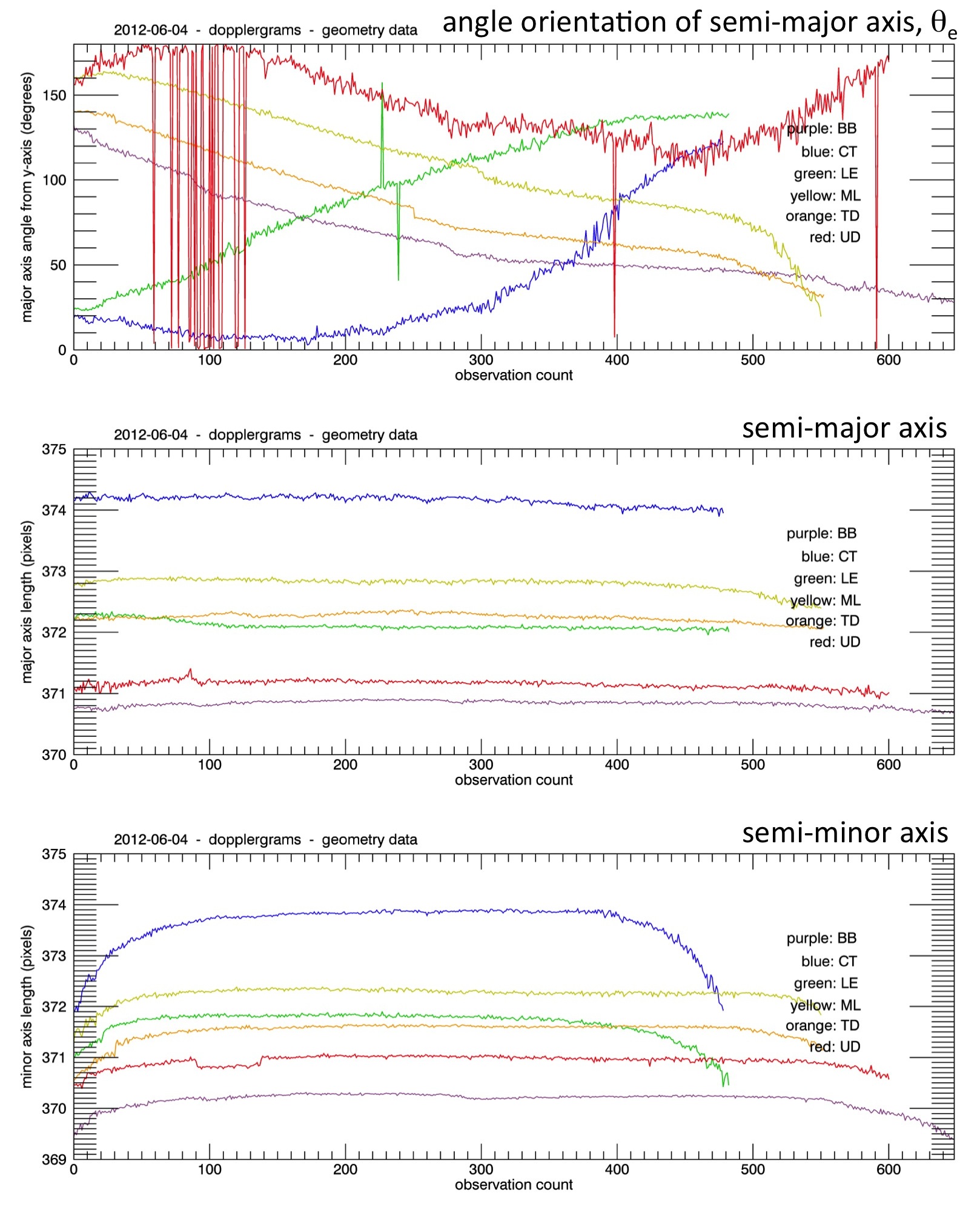}
     \caption[Solar-image ellipticity values through a day]{Plots of several ellipticity values available as fit parameters within the fits headers.  The orientation of the solar-ellipse (top panel) varies greatly throughout the day, whereas the semi-major radius (middle panel) remains relatively constant.  However, depending on the site and the time of year, the ellipticity as depicted by semi-minor radius (bottom panel) varies noticeably between noon-time and morning/evening due to refraction effects.}
     \label{FIG_Dataellipticity}
     \end{center}
     \end{figure}

Figure~\ref{FIG_Dataellipticity}
presents these three fit parameters across a full-day's worth of observations.  From this, we see that the orientation of the ellipse varies {\em relatively} smoothly throughout an observing day, and that the degree of ellipticity for each observation is very much a function of the time of day and how close the sun is to the local horizon, with a maximum pixel difference between semi-major and semi-minor axes on the order of two pixels.

     \subsection[\textcolor{blue}{Elliptical Solar Images}]{\textcolor{blue}{Elliptical Solar Images}}
     \label{Data_Ellipses}

As described above, the ellipticity in the GONG site images is quite small, adding up to a few pixels limb-position difference at the most.  However, as discussed in the results of \S\ref{Results_DopHalf_Filtered}, the effect is still strong enough to compete noticeably with the level of accuracy needed for matching against the time-of-day curves output by OFFSET by distorting the perceived location of the axis-of-rotation.

\begin{figure}
     \begin{center}
     \includegraphics[scale=0.18]{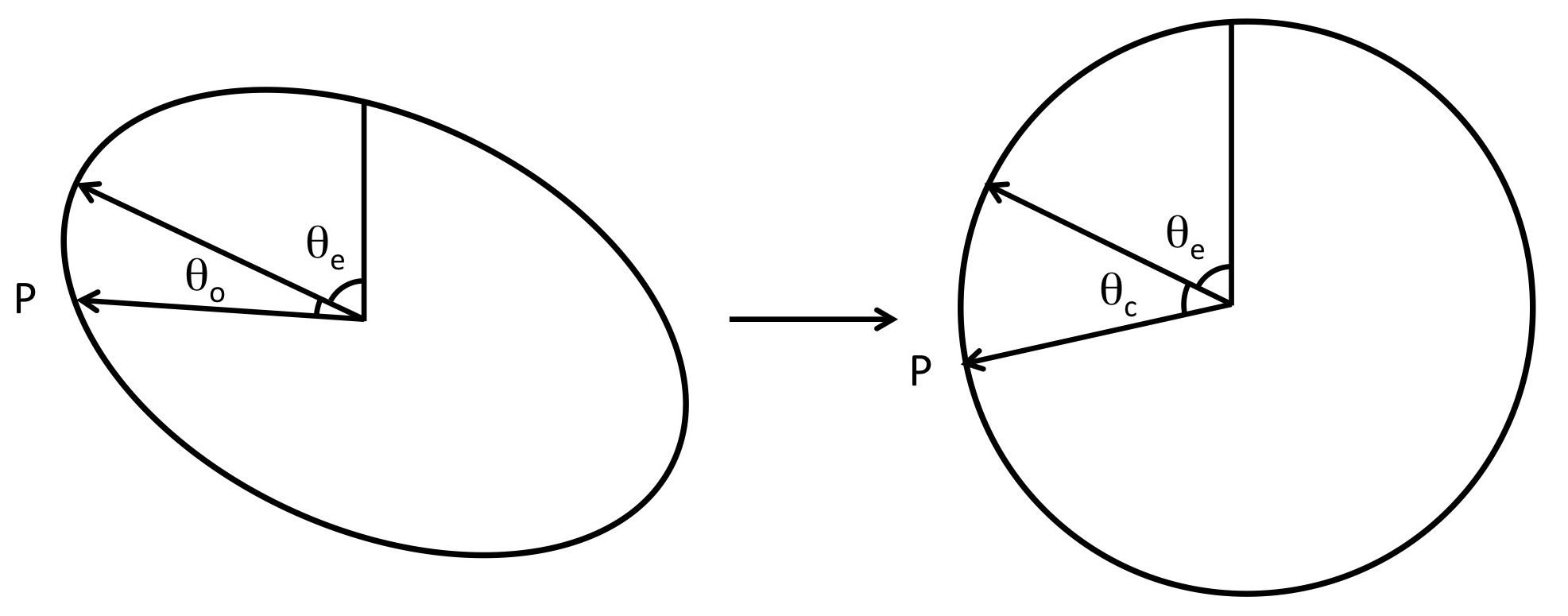}
     \caption[Ellipse-deprojection correction]{Cartoon of the ellipse-deprojection correction.}
     \label{FIG_DataEllipseCorrect}
     \end{center}
     \end{figure}

A method used in this report to try to account for this distortion is to de-project the vector of the observed rotation axis from the elliptical into a circular frame.  This method assumes that the solar image has merely been squished along the direction of the semi-minor axis (or stretched along the semi-major axis).  Therefore, a point (in this case the north- (or south-)pole) defined by some angle relative to the semi-major-axis angle, $\theta_e$, can be shifted back into proper alignment by restoring the vector components into an $s_\mathrm{major} = s_\mathrm{minor}$ reference frame, as depicted in Figure~\ref{FIG_DataEllipseCorrect}.

In order to make this correction, we first solve for the axis-angle relative to the ellipse angle as
\begin{equation}
   \theta_\circ
=
   \theta_{P,\mathrm{obs}} - \theta_e
   \, ,
\end{equation}
where $\theta_{P,\mathrm{obs}}$ is the initially observed P-angle in the image.
Next, we find the components of $\theta_\circ$:
\begin{eqnarray}
   a
\nonumber
&=&
   \cos \theta_\circ
   \,,
\\
   b
&=&
   \sin \theta_\circ
   \,.
\end{eqnarray}
These components are then re-scaled into a circular reference frame, using:
\begin{eqnarray}
   a'
&=&
   a / s_\mathrm{major}
   \,,
\nonumber
\\
   b'
&=&
   b / s_\mathrm{minor}
   \,.
\end{eqnarray}
Finally, the corrected relative angle is given by:
\begin{equation}
   \theta_c
=
   \arctan \left( \frac{b'}{a'} \right)
   \,,
\end{equation}
providing the corrected P-angle:
\begin{equation}
   \theta_P
=
   \theta_e + \theta_c
   \,.
\end{equation}

Unfortunately, it is not always clear {\em when} this correction needs to be applied.  The method for angle-bin sampling within a {\em circular} reference frame used in the RingPhase algorithm suggests that ellipse de-projection {\em should} be used to correct the RingPhase results.  However, in \S\ref{Results_DopHalf_Filtered} we see that there is some evidence to suggest that DopHalf's method of sampling within the elliptical image-disk {\em may already} account for this correction (though this is not certain).  In the ideal situation, images would first be re-registered as circular, rendering this correction unnecessary.

     \subsection[\textcolor{blue}{Solar B0-angle}]{\textcolor{blue}{Solar B0-angle}}
     \label{Data_B0}

\begin{figure}
     \begin{center}
     \includegraphics[scale=0.18]{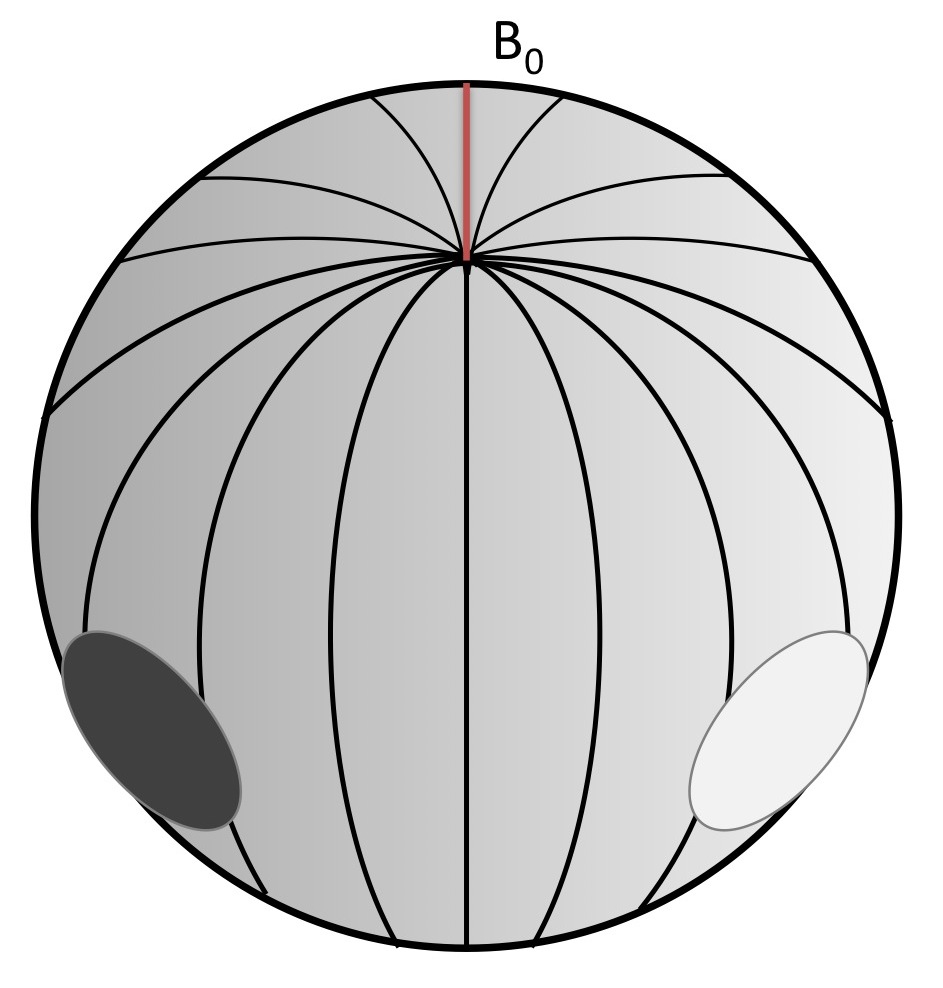}
     \caption[Solar $B_0$-angle]{Cartoon of the solar $B_0$-angle.}
     \label{FIG_DataB0}
     \end{center}
     \end{figure}

Finally, there is also the matter of the physical orientation of the solar-rotational axis relative to the viewing geometry.  Because the rotational axis of the sun is inclined roughly $7^\circ$ to the plane of the ecliptic \cite{BeckGiles2005}, the line of the equator as plotted across a solar image varies throughout the year, so that at some times the equator cuts directly through the center of the solar disk, and at other times it moves north or south, with some fraction of the opposite pole becoming visible instead, as shown in Figure~\ref{FIG_DataB0}.
The angle between the rotational-pole and the line-of-observation is given by the B0-angle.

For minimum/maximum B0, the peak near-equator values in rotational velocity become noticeably shifted north or south in the Dopplergram image.  While the effect of this peak shift on P-angles computed using the DopHalf method is not clear and may be minor, this poses obvious problems for calculations using the RingPhase method, since the rotational signature is no longer well-represented by a single-period sinusoid moving angularly around the solar limb.  Therefore, for the testing presented in this report, we use observation sets taken primarily in early June when the B0-angle is near 0-degrees.

\section[\textcolor{blue}{Results}]{\textcolor{blue}{Results}}
\label{Results}

In the following sub-sections, we describe the testing that we did on the DopHalf (\S\ref{Results_DopHalf}) and RingPhase (\S\ref{Results_RingPhase}) algorithms and the results we have acquired to date.  In the DopHalf section, tests are focused  primarily on possible ways to manipulate the Dopplergram image before using Equations~\ref{EQ_AlgoDopHalfP} and \ref{EQ_AlgoDopHalfVmean} to calculate $\theta_P$.  In the RingPhase section, there is a greater focus on the global {\em radial-structure} dependence of the outputted $\theta_P$ results, making comparisons also with radially dependent DopHalf results.  Finally, in \S\ref{Results_VelGrad}, we consider a more general comparison between these Dopplergram-based algorithms and the time-of-day curves supplied by OFFSET, and present some results concerning a toy model for the way site Dopplergrams may or may not be sufficiently calibrated across the image frame.

     \subsection[\textcolor{blue}{DopHalf Performance}]{\textcolor{blue}{DopHalf Performance}}
     \label{Results_DopHalf}

The DopHalf algorithm is based on the algorithm described in a study using MDI data \cite{MDIDopHalf} and is fundamentally a very quick and simple way to evaluate the Dopplergram data.  In the following sub-sections, we will first present the results that we have acquired using DopHalf on raw Dopplergram data (\S\ref{Results_DopHalf_Raw}).  Next, we will discuss the Dopplergram filtering methods explored and the general results of applying those to the image before running the DopHalf calculation (\S\ref{Results_DopHalf_Filtered}).  Finally, we compare the different explored DopHalf approaches using a set of HMI test images modified to different rotation orientations and look more closely at the effects of input-image orientation on the DopHalf results (\S\ref{Results_DopHalf_HMIrot}).

          \subsubsection[\textcolor{blue}{Using Raw Dopplergrams}]{\textcolor{blue}{Using Raw Dopplergrams}}
          \label{Results_DopHalf_Raw}

\begin{figure}
     \begin{center}
     \includegraphics[scale=0.1]{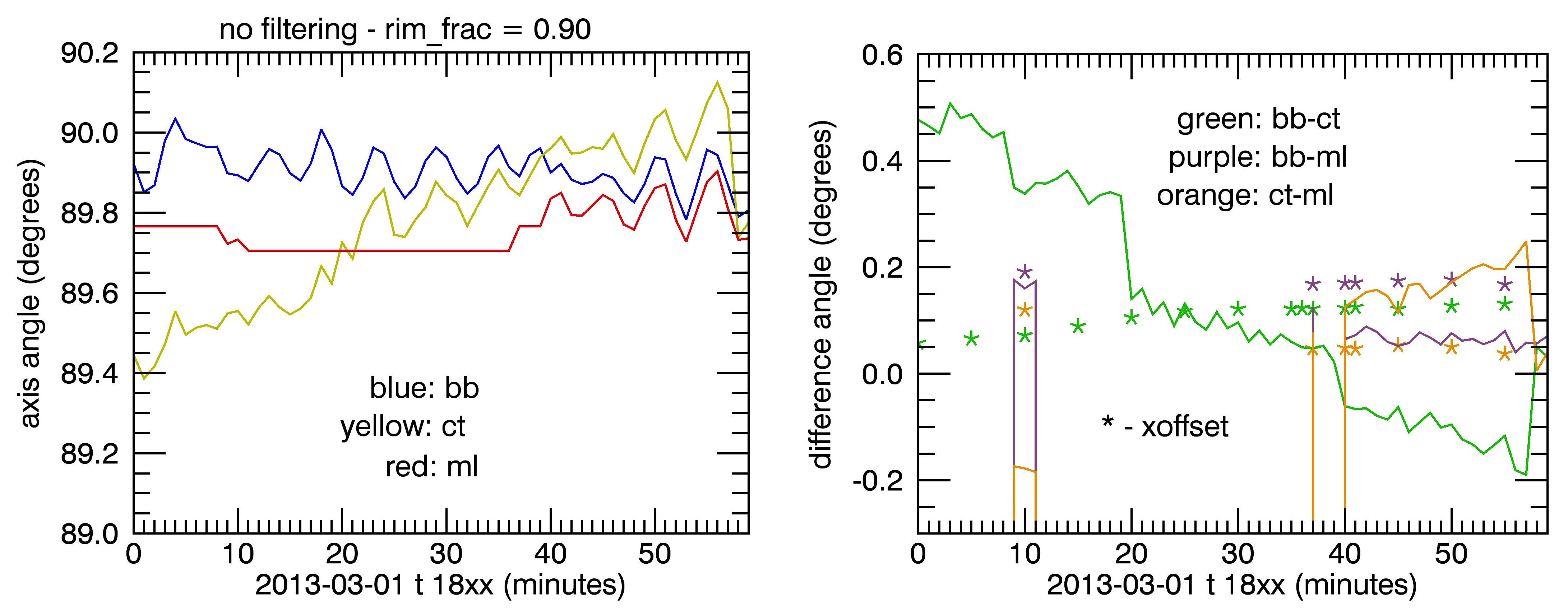}
     \caption[Early DopHalf results compared to xoffset]{Early DopHalf results (left panel) as well as site-to-site DopHalf-difference angles compared to the xoffset results (right panel).  Notice that the minute-to-minute structure of the DopHalf results from different sites mirrors each other fairly well at some times, e.g., ~45-60 minutes after UT 18:00.  However, there does not appear to be a clear correlation between the DopHalf-difference angles and the results output by xoffset.  (Also note that this data set represents a serendipitously tight set of P-angle results from DopHalf.)  'rim\_frac = 0.90' means that DopHalf sampled the solar image out to a radius of 90\% the limb-fit radius.}
     \label{FIG_Results_DopHalf_Raw_vsxoffset}
     \end{center}
     \end{figure}

In Figure~\ref{FIG_Results_DopHalf_Raw_vsxoffset},
we present the results of one of our earliest tests, applying the DopHalf algorithm to a short time-series of concurrent Big Bear, Cerro Tololo, and Mauna Loa site data.  The P-angles thus computed are presented in the left panel and show that the minute-to-minute variations in the signal appear reasonably noisy.  However, as explained in the SOHO/MDI study \cite{MDIDopHalf}, much of that 'noise' has a basis in the actual physical state of the sun at the times of observation.  Therefore, there are at least some points of consistency in the curves from one site to another.  This provides hope that perhaps those variations caused by things like p-mode oscillations are not {\em too} detrimental to making site-to-site P-angle comparisons using DopHalf.  

However, the right-hand panel of Figure~\ref{FIG_Results_DopHalf_Raw_vsxoffset} shows the computed difference angles between the site-image sets compared to computed xoffset difference-angle results.  The comparison is not strong.  And although this is early data produced by still-buggy DopHalf code, this particular time of day and set of site observations and results will later be shown to be serendipitously {\em well-aligned} to the expected $\sim 90^\circ$ base orientation.

\begin{figure}
     \begin{center}
     \includegraphics[scale=0.30]{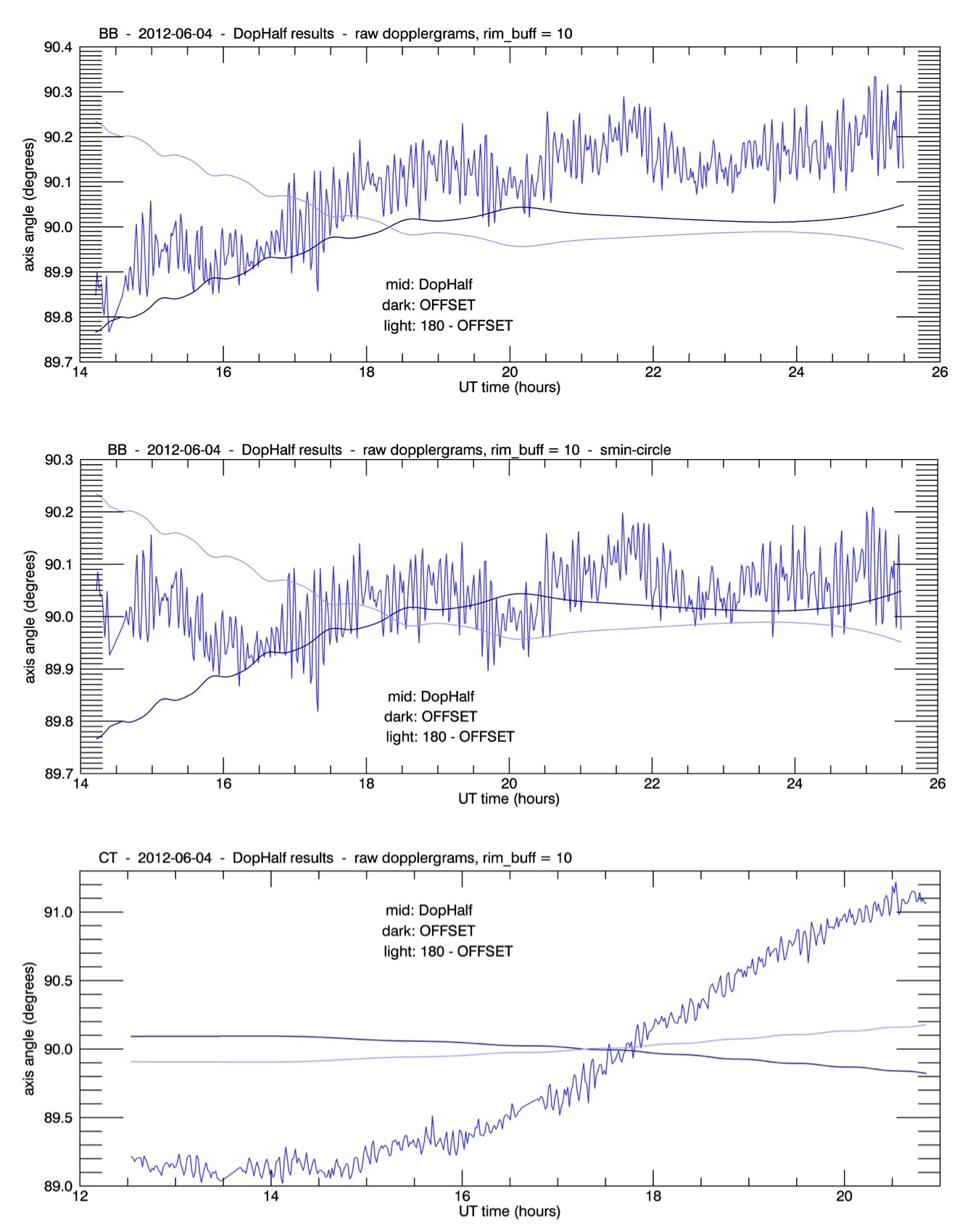}
     \caption[DopHalf results using raw Dopplergrams versus OFFSET]{DopHalf results using raw GONG-site Dopplergrams from Big Bear and Cerro Tololo as a function of time-of-day; compared against the OFFSET values reported in the input file headers.  {\bf Note that because of the double effects of GONG site images being flipped east-to-west and the fact that OFFSET angles are measured clock-wise rather than counter-clockwise from the +y-axis, the results from DopHalf should always be compared to the 180$^\circ$-OFFSET curves.}  The top and bottom panels present Big-Bear- and Cerro-Tololo-DopHalf results using sampling that adheres to the elliptical-fit shape of the solar image, while the middle panel presents results of sampling the Big Bear images on a circular disk only, defined by a radius equal to the semi-minor axis value.  'rim\_buff = 10' means that a buffer of 10 pixels from the limb were excluded from the hemispheric-mean calculations for DopHalf.}
     \label{FIG_Results_DopHalf_Raw_vsOFFSET}
     \end{center}
     \end{figure}

Next, in Figure~\ref{FIG_Results_DopHalf_Raw_vsOFFSET},
we present an example of the DopHalf code run on raw Dopplergram data for a full observing-day's worth of site data, in this case taken from Big Bear and Cerro Tololo.  The DopHalf curves are presented compared with the input images' computed OFFSET values.  Now we can see that on top of the minute-to-minute variations, there are longer-period variations to the P-angle curve that make aligning this set of DopHalf results to the OFFSET curves problematic.  Further, while the Big-Bear curve at least remains within $0.5^\circ$ of the OFFSET curve {\em including} noise/oscillations, the Cerro-Tololo raw-Dopplergram results deviate from the OFFSET curve by more than a full degree.

There are two other issues presented in Figure~\ref{FIG_Results_DopHalf_Raw_vsOFFSET} to note.  First is the fact that various orientation issues mean that for correct comparison of the DopHalf results we must compare to $180^\circ$ - OFFSET, rather than to the OFFSET values directly.  Later in this report, for experiments comparing DopHalf results to RingPhase results, we instead switch to presenting the DopHalf results as $180^\circ$ - DopHalf-code-output.  Second, because the results in the middle panel of Figure~\ref{FIG_Results_DopHalf_Raw_vsOFFSET} were computed using a circular-disk sampling, comparison of the top and middle panels provides a first-order assessment of the affect of the image-ellipticity on the DopHalf calculation.  While the effect is not on the same order as the Cerro-Tololo deviation from the OFFSET curve, it still has a noticeable impact on the full-day comparison of the Big-Bear results to the Big-Bear OFFSET curve.

          \subsubsection[\textcolor{blue}{Using Filtered Dopplergrams}]{\textcolor{blue}{Using Filtered Dopplergrams}}
          \label{Results_DopHalf_Filtered}

\begin{figure}
     \begin{center}
     \includegraphics[scale=0.2]{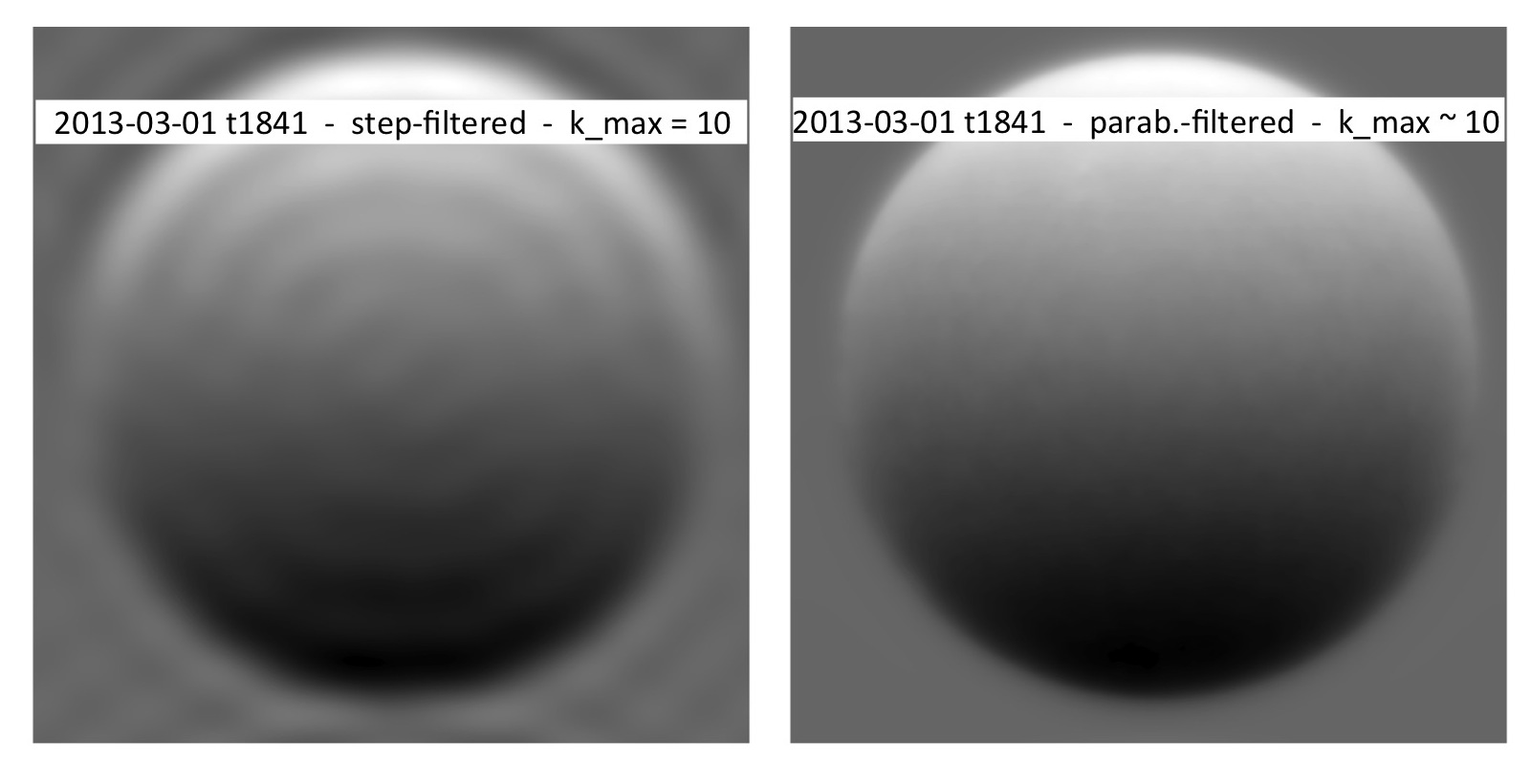}
     \caption[Solar Dopplergrams filtered according to wavelength]{Images of GONG-site Dopplergrams from Big Bear after low-pass filtering to retain only those FFT components with frequencies less than about 10 periods across the image.  The left panel displays the results of filtering with a sharp cut-off after 10 periods.  The right panel displays the results of using a parabolically tapered filter.}
     \label{FIG_Results_DopHalf_Filtered_kimages}
     \end{center}
     \end{figure}

In this section, we present possible methods for filtering the GONG-site Dopplergrams in the hopes of suppressing some of the non-rotational signals contained within these images.  The first, most obvious filter to try is one that clips out the high-spatial-frequency velocity signals.  Figure~\ref{FIG_Results_DopHalf_Filtered_kimages}
presents images of two such filtered Dopplergrams.  To apply this filter, we first compute the 2D FFT of the Dopplergram, then mask out the high-frequency FFT components, before inverting the FFT to produce the filtered Dopplergram image.  In the left panel, the mask has been applied as a step function, which produces a clear 'ringing' effect in the final image.  In the right panel, we have applied a parabolic-function-mask to suppress the high-frequencies, but less abruptly.  This was the first filter tested, early in the DopHalf code development, and no clear improvement was seen in the comparison to xoffset data.  Therefore, testing proceeded to a new filter.

\begin{figure}
     \begin{center}
     \includegraphics[scale=0.2]{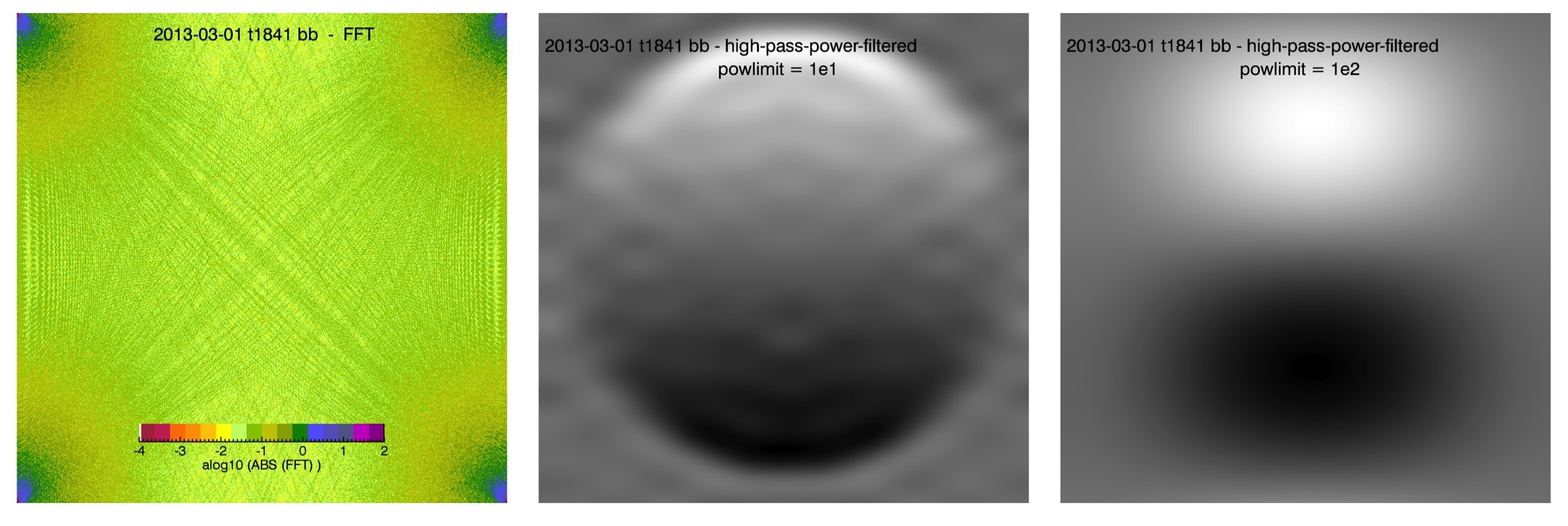}
     \caption[Solar Dopplergrams filtered according to power]{Images of GONG-site Dopplergrams from Big Bear after low-pass filtering to retain only those FFT components with power above a certain threshold.  The left-most panel displays the power in the computed FFT for the image.  The other two images display results of filtering for power $> 10$ and power $> 100$.}
     \label{FIG_Results_DopHalf_Filtered_pimages}
     \end{center}
     \end{figure}

\begin{figure}
     \begin{center}
     \includegraphics[scale=0.2]{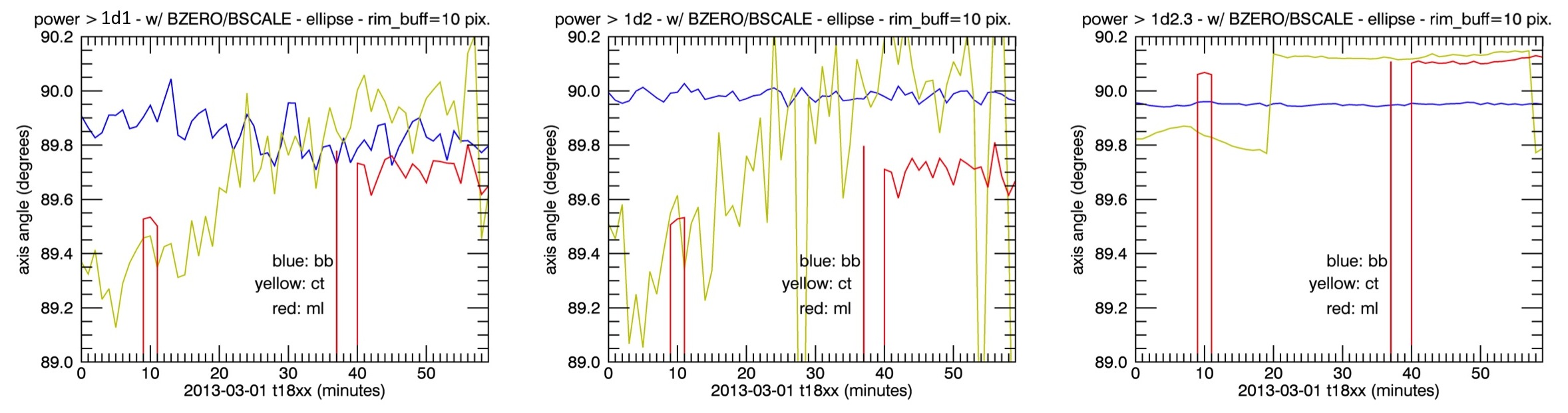}
     \caption[Early DopHalf results for high-power-filtered Dopplergrams]{Early DopHalf results using Dopplergrams filtered to retain only those FFT components with power above some threshold.  Left panel: power $> 10$; middle panel: power $> 100$; right panel: power $> 10^{2.3}$.}
     \label{FIG_Results_DopHalf_Filtered_pDopHalfs}
     \end{center}
     \end{figure}

The next filter tried was designed to select for the highest-power components of the image FFT result, and some filtered-image examples are presented in Figure~\ref{FIG_Results_DopHalf_Filtered_pimages}
with an image of the FFT power-spectrum presented in the left-most panel.  Rather than ringing, the effects of this filter appear to be a sort of dimpling in the image, at least if only moderately low-power components are excluded.  As with the first filter, DopHalf tests of the high-power-pass filter appeared discouraging, with noise in the signal {\em increasing} for increasing power-filter cutoffs, as shown in the first-two panels of Figure~\ref{FIG_Results_DopHalf_Filtered_pDopHalfs}.

\begin{figure}
     \begin{center}
     \includegraphics[scale=0.18]{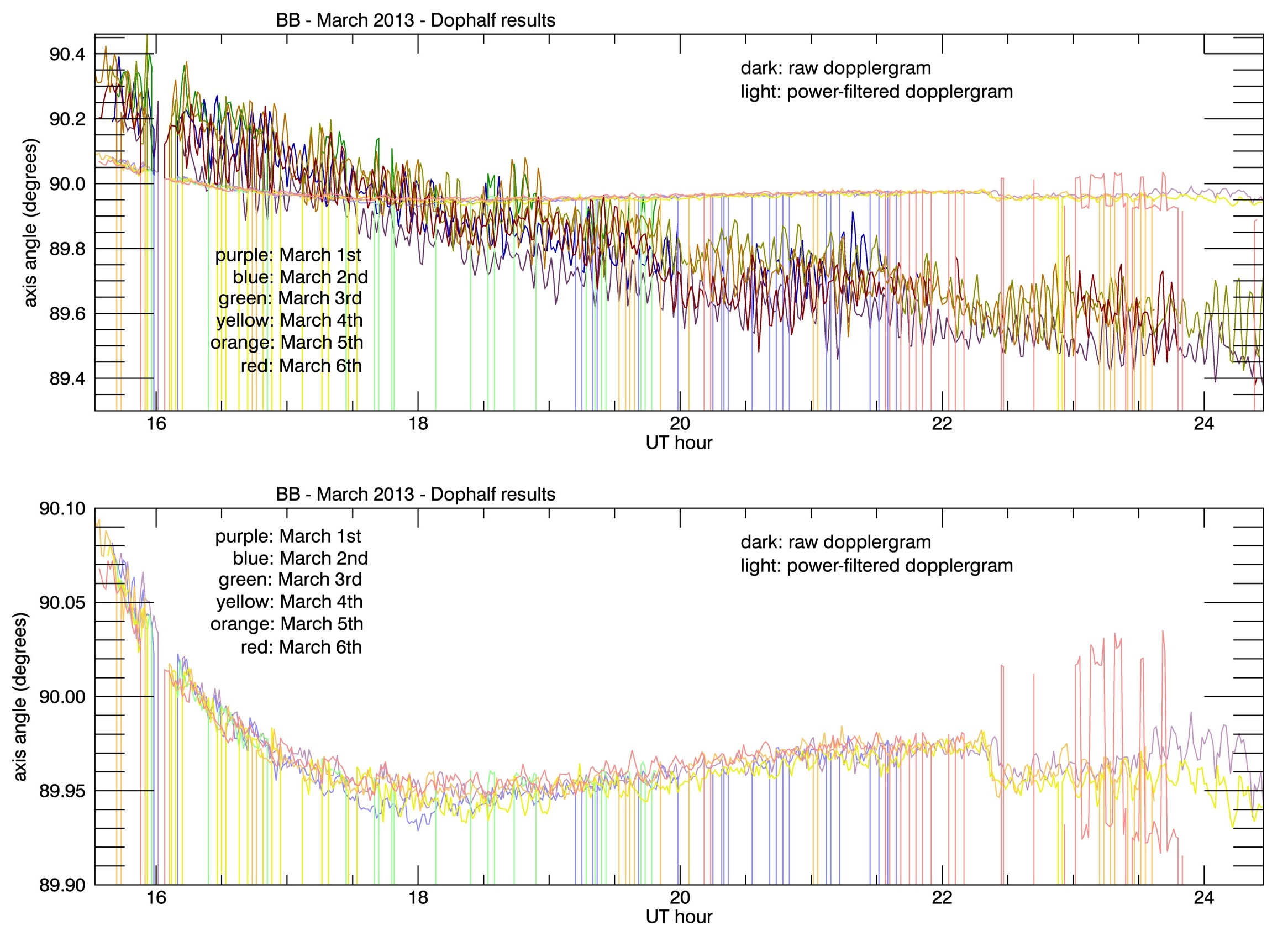}
     \caption[Raw and top-power-filtered DopHalf results over several days]{Results of DopHalf applied to Big Bear-site data for several consecutive days using raw Dopplergrams (top panel only) compared to using top-power-filtered Dopplergrams.  While the filtered-Dopplergram results tend to appear much more stable, an exception occurs on March 6th around UT 23:00 and is due to the fact that these results were computed using a minimum-power-cutoff of $10^{2.3}$, rather than by directly selecting the single-highest-power FFT component.}
     \label{FIG_Results_DopHalf_Filtered_toppDopHalfs_days}
     \end{center}
     \end{figure}

However, for a {\em very} high power cutoff of $10^{2.3}$, the DopHalf results in the sample set appear suddenly quiet and well behaved.  Testing determined that this result occurred if the filter mask selected for only the {\em single} highest-power signal in the FFT computation for an image.  Furthermore, the DopHalf results using this filter appear quite self-consistent from one site day to another, as shown in Figure~\ref{FIG_Results_DopHalf_Filtered_toppDopHalfs_days},
and that the curve produced across a full observing day remains much nearer the nominal $90^\circ$ than when using the raw Dopplergrams.

     \begin{figure}
     \begin{center}
     \includegraphics[scale=0.18]{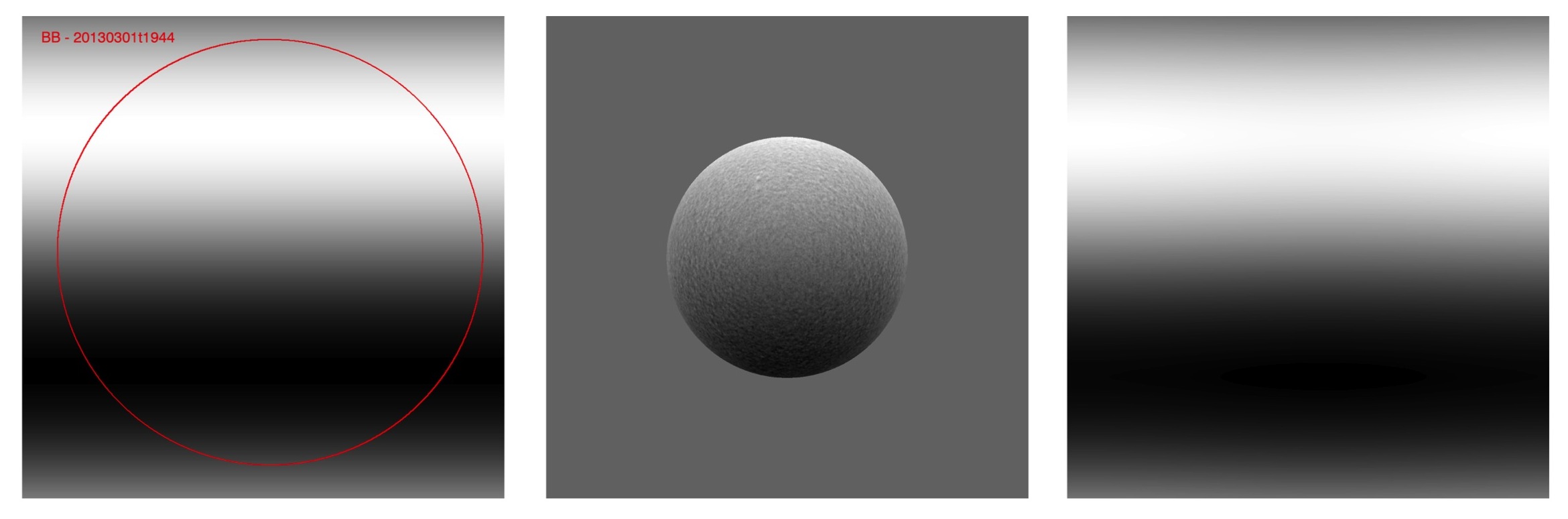}
     \caption[Images of different Dopplergram-filter results]{Comparison of Dopplergrams filtered under different specifications.  Left panel: using the top-power FFT component, i.e., a y-wave-only component.  Middle and Right panels: using the FFTwide filtering scheme: first (middle panel) the Dopplergram is copied onto a larger image array twice the width of the solar image and the background set to the solar-image-mean value; then the filtered result (right panel) uses {\em both} the longest x and y-direction wave components.}
     \label{FIG_Results_DopHalf_Filtered_topp_v_fftwide_images}
     \end{center}
     \end{figure}

{\em However}, further investigation into the top-power-pass filter determined that it is {\em not} viable for determining the P-angle of a given observation.  Because all of the site observations are oriented with the rotational axis nearly parallel to the x-axis of the image, the highest-power FFT component is in fact the longest-wavelength y-direction component, in every image.  The left-most image displayed in Figure~\ref{FIG_Results_DopHalf_Filtered_topp_v_fftwide_images}
shows an example of a top-power--filtered Dopplergram, with an overlaid outline of the fit solar-limb ellipse for that observation.  This image contains only a single gradient in the y-direction.  The only mechanism by which DopHalf does {\em not} compute a P-angle of always $90^\circ$ for this filtered image is the ellipse-bounded sample space for each hemispheric-mean calculation.  This derives a 'P-angle' somewhat offset from $90^\circ$, which varies relatively smoothly throughout the day because the orientation of the solar ellipse with respect to the telescope does so, as was seen in Figure~\ref{FIG_Dataellipticity}.

\begin{figure}
     \begin{center}
     \includegraphics[scale=0.24]{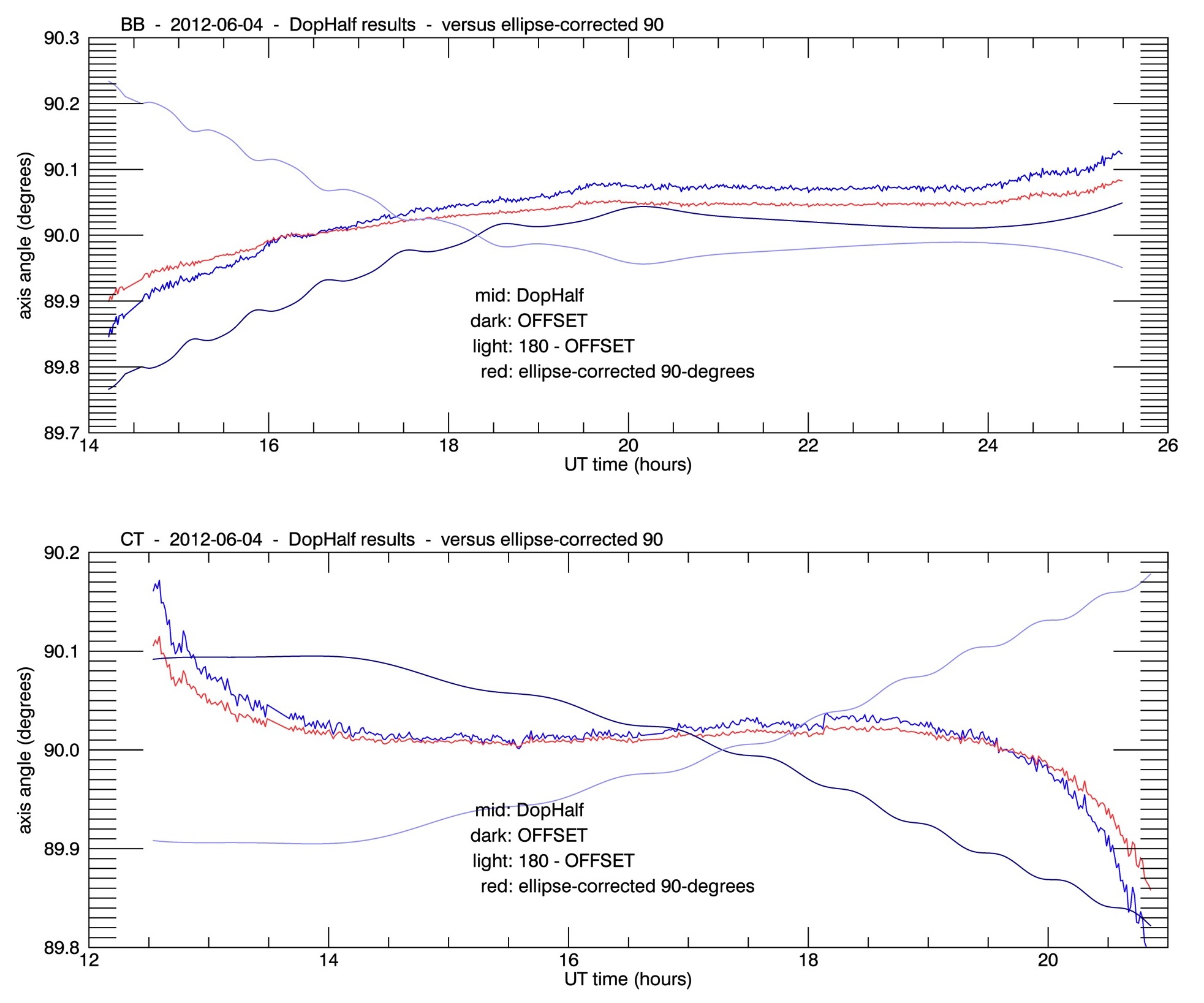}
     \caption[Top-power-filtered DopHalf results versus the ellipse-deprojection-correction]{Results of DopHalf (on top-power--filtered Dopplergrams) applied to Big Bear (top panel) and Cerro Tololo (bottom panel) site data compared against a curve (red)  computed using the ellipse-de-projection correction (see \S\ref{Data_Ellipses}).  The de-projection correction was applied to a fixed 'observed' P-angle value of $90^\circ$ for each of the input-solar-image geometries.}
     \label{FIG_Results_DopHalf_Filtered_toppDopHalfs_vEangle}
     \end{center}
     \end{figure}

To demonstrate that the above statement is true, we present Figure~\ref{FIG_Results_DopHalf_Filtered_toppDopHalfs_vEangle}.
This figure presents the DopHalf full-day curves for two sites computed using the top-power-filtered / y-wave-only Dopplergrams along-side a second curve.  The only input into this second curve (shown in red) comes from the ellipse-shape and orientation values reported in the input headers which were then fed into the ellipse-de-projection-correction algorithm (\S\ref{Data_Ellipses}) assuming an 'observed' P-angle of exactly $90^\circ$ for every observation.  While the two curves are not exact matches, they are too closely aligned for the described relationship not to exist.

\begin{figure}
     \begin{center}
     \includegraphics[scale=0.065]{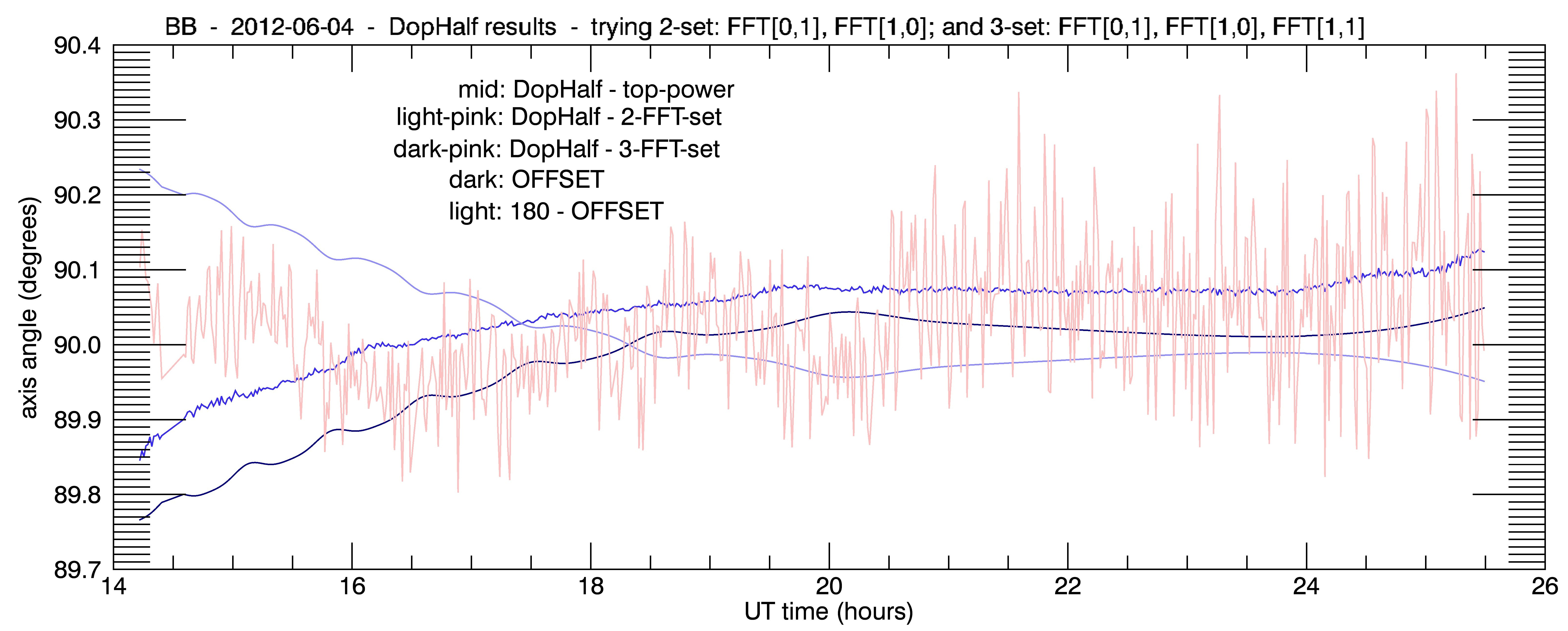}
     \caption[DopHalf results using multiple longest-wavelength FFT components]{Results of DopHalf applied to Big Bear site data using Dopplergrams filtered using a couple of the longest-wavelength FFT components.  The mid-blue curve corresponds to the top-power filtering which is equivalent to using {\em only} the longest y-direction component.  The light-pink curve corresponds to using both the longest x and y-direction components.  Another curve also exists off the scale of this figure.  It was computed by including also the longest x+y (FFT[1,1]) component during the image filtering.}
     \label{FIG_Results_DopHalf_Filtered_xylongestresults}
     \end{center}
     \end{figure}

As a result of the above findings, we tested a final set of Dopplergram filters.  For these filters, we select two specific FFT components for each observation image: the y {\em and} x longest-wavelenth FFT components.  The results of one full-day set can be seen in Figure~\ref{FIG_Results_DopHalf_Filtered_xylongestresults}
plotted against the top-power-pass-filtered results as a reference.  These results (pink curve) are strongly reminiscent of the raw-Dopplergram results (Figure~\ref{FIG_Results_DopHalf_Raw_vsOFFSET}), though they are also noisier.

Never-the-less, there is some chance (depending on the setup of the input image) that the FFT components themselves may contain further information on the orientation of the Dopplergram.
Therefore, we have also tested a variation on this FFTxy filter that applies this filter and the DopHalf computation to a copy of the input Dopplergram that has been remapped onto the center of an image array precisely twice the width of the sun-image mean radius.  An example of this starting Dopplergram and the resulting FFTwide-filtered image are shown in the middle and right-most panels of Figure~\ref{FIG_Results_DopHalf_Filtered_topp_v_fftwide_images}.  Note that the filtered result appears very similar to the earlier y-only-filtered image (leftmost panel).  In this example, the y-component has an amplitude nearly 49 times greater than the x-component, with a phase angle of $89.004^\circ$ to the x-component's $-11.364^\circ$.  Further results on the use of this FFTwide filtering method are presented in the next sub-section.

          \subsubsection[\textcolor{blue}{Comparison to HMI-test and Rotated Images}]{\textcolor{blue}{Comparison to HMI-test and Rotated Images}}
          \label{Results_DopHalf_HMIrot}

In this section, we compare the DopHalf calculation and the different viable Dopplergram-filter options using a set of test images and examine the effects of image orientation on the results.

The test images were prepared by using HMI Dopplergram data scaled to the image-size of GONG-site data.  The data all use the same circular-solar-image Dopplergram positioned at the same (x,y) center point but with the P-angle axis rotated by some known amount.  The results of this test set are presented in Table~\ref{TAB_Results_DopHalf_HMIrot_values}, 
\begin{table}
\caption{HMI-test Rotated-images DopHalf Results}
\centering
\begin{tabular}{c | c c | c c | c c}
	\hline \hline
	Rotation & 
	   raw & $\delta$-raw & 
	   xy-longest & $\delta$-xy-l. &
	   FFTwide & $\delta$-FFTw.
	   \\ [0.5ex]
	\hline
	44	& 45.626	& 1.63	& 45.400	& 1.40	& 45.454	& 1.45	\\
	44.5	& 46.104	& 1.60	& 45.863	& 1.36	& 45.951	& 1.45	\\
	45	& 46.570	& 1.57	& 46.308	& 1.31	& 46.437	& 1.44	\\
	45.5	& 47.023	& 1.52	& 46.740	& 1.24	& 46.904	& 1.40	\\
	46	& 47.512	& 1.51	& 47.211	& 1.21	& 47.405	& 1.41	\\ [1ex]
	89.5	& 91.193	& 1.69	& 90.240	& 0.74	& 90.811	& 1.31	\\
	90	& 91.757	& 1.76	& 90.660	& 0.66	& 91.294	& 1.29	\\
	90.5	& 92.253	& 1.75	& 91.307	& 0.81	& 91.826	& 1.33	\\
	91	& 92.770	& 1.77	& 91.830	& 0.83	& 92.329	& 1.33	\\ [1ex]
	\hline
\end{tabular}
\label{TAB_Results_DopHalf_HMIrot_values} 
\end{table}
showing the DopHalf results using raw Dopplergrams, Dopplergrams filtered to include the longest-x\&y FFT components only, and Dopplergrams run through the FFTwide remapping/filtering scheme.  The results are reported in degrees.  The second column for each set presents the rounded difference angle between the results and the known image rotation.

One point of note is that this is for a single observation, rather than a time-series of observations.  Therefore, there is likely to be some intrinsic DopHalf offset due to the solar p-mode oscillations that produce the minute-to-minute variations seen in the DopHalf results previously.  These HMI-test-image results suggest that DopHalf produces a moderately consistent offset to the expected rotation angle.  However, the computed offset varies noticeably between the raw-Dopplergram and filtered-Dopplergram sets, and for the simple FFTxy-filtereing set there is a {\em clear} difference in the computed offsets for the image orientations near $90^\circ$ versus near $45^\circ$.

\begin{figure}
     \begin{center}
     \includegraphics[scale=0.18]{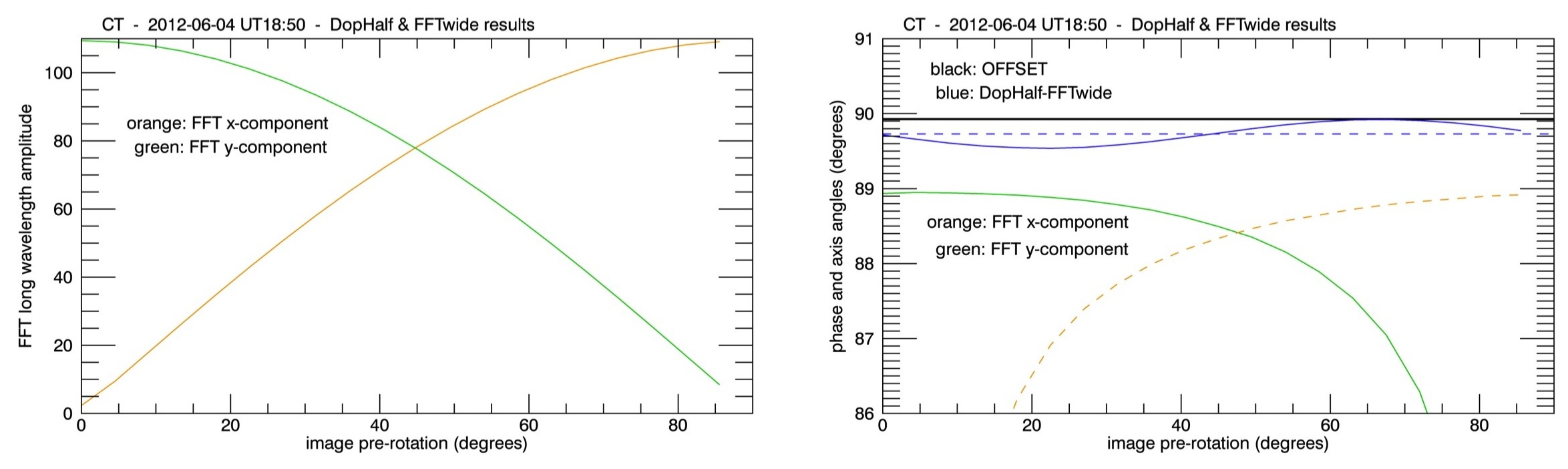}
     \caption[Geometry results of rotating the Dopplergram before performing FFTwide filtering]{Geometry results of rotating the Dopplergram through a range of angles before performing FFTwide filtering and computing DopHalf results.  The left panel shows that the amplitude of the longest-wavelength x- and y-FFT components equalize very close to $45^\circ$ rotation.  The right panel plots both the P-angle results (blue and black curves) and the absolute-value phase-angles of the two employed FFT components (orange and green curves).  A phase-angle of $90^\circ$ would suggest an FFT-component that is centered, both on the image and on the solar Dopplergram.  The dashed-blue curve represents the mean of the FFTwide-DopHalf results.}
     \label{FIG_Results_DopHalf_rot_FFTwidestuff}
     \end{center}
     \end{figure}

Next, we take a look at the effect of rotating a GONG site image as an added step in the DopHalf processing routine.  In the case of the FFTwide filtering, this is particularly of interest to see if there might be some orientation where the phase angles of the x and y FFT components are both near $\pm 90^\circ$, which would suggest a solar-image-centered filtering.  In Figure~\ref{FIG_Results_DopHalf_rot_FFTwidestuff},
we present the FFT component attributes for a Dopplergram image rotated through $0^\circ$-$90^\circ$.  Not surprisingly, the amplitudes of the two components (left panel) vary in supremacy depending on the image orientation, with equal amplitudes occurring close to or at $45^\circ$.  In terms of the component centering on the image, the phase-angle curves of the two components (right panel) also cross paths near $45^\circ$.  However, the drop-off away from $\pm 90^\circ$ is steeper away from this point and the intersection occurs neither right {\em at} $45^\circ$ nor right at the point where the computed P-angle curve (which does vary noticeably (and sinusoidally) throughout the rotation) intersects its mean.

\begin{figure}[t]
     \begin{center}
     \includegraphics[scale=0.18]{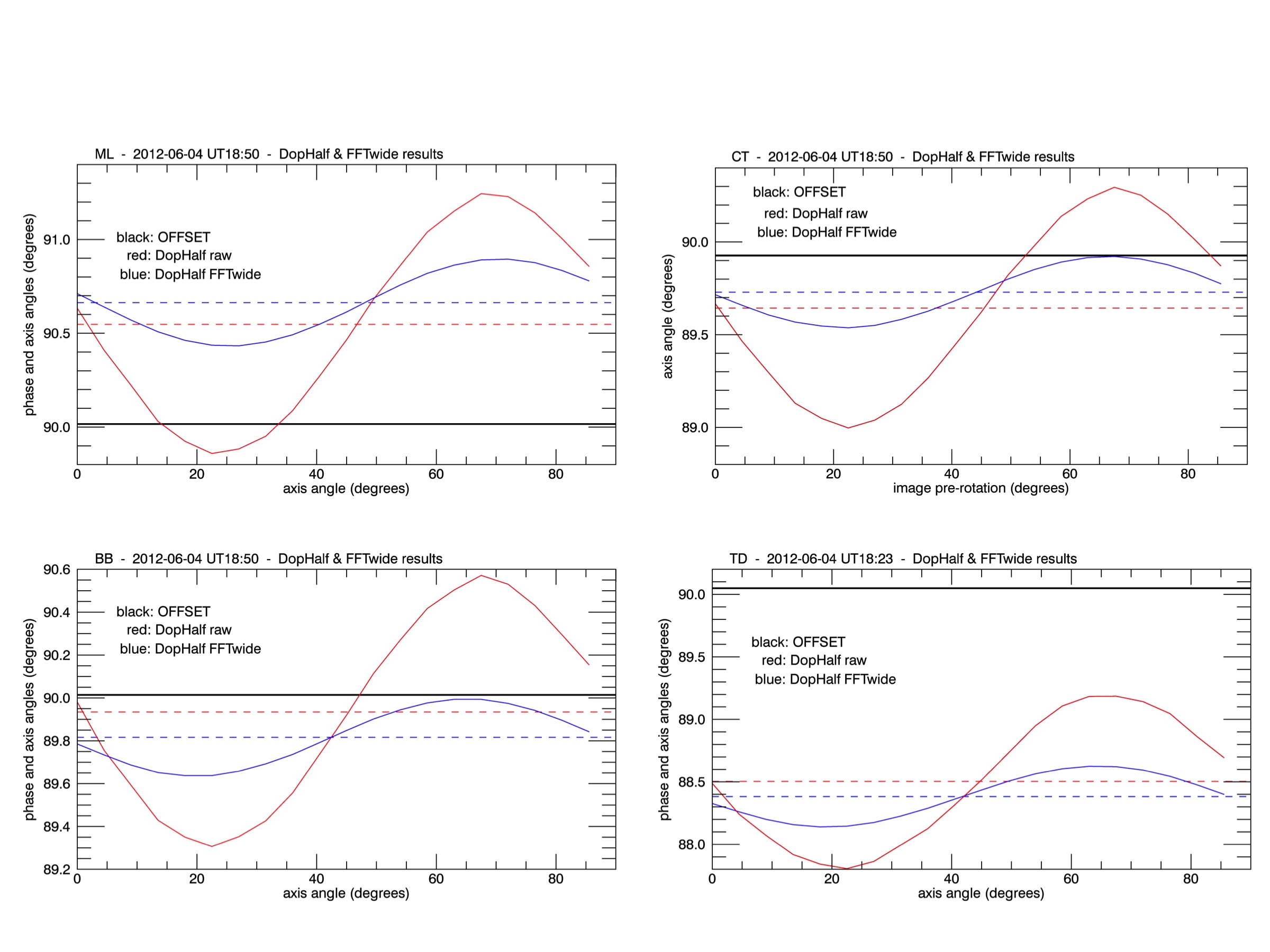}
     \caption[DopHalf example results from rotating the Dopplergrams for four sites]{Example results from DopHalf using both raw and FFTwide-filtered Dopplergrams for images from four sites that have been rotated through a range of $90^\circ$.  Note that the Cerro Tololo curve only serendipitously matches the OFFSET value at the peak of the FFTwide-results curve.  Otherwise, the means of the two DopHalf methods do not agree, even in terms of which curve is closest to the OFFSET value.  
     The mean values occur near (but not precisely at) $0^\circ$, $45^\circ$, and $90^\circ$.}
     \label{FIG_Results_DopHalf_rot_4sitescurves}
     \end{center}
     \end{figure}

Taking a closer look at the DopHalf-P-angle curve generated by rotating the image, Figure~\ref{FIG_Results_DopHalf_rot_4sitescurves} 
presents results using raw and FFTwide-filtered Dopplergrams from four sites.  The curves are sinusoidal, covering a full period in a $90^\circ$-image-rotation.  The DopHalf curves using raw Dopplergrams have a distinctly larger amplitude, covering a total of about $1.5^\circ$ in P-angle space.  The FFTwide filtered curves cover about a third of that.  The means of the two curves are offset by some variable amount from one another, and not all of the rotated-DopHalf curves have an intersection point with the reported OFFSET value.  One other point of note is that the mean values of the curves occur near $0^\circ$, $45^\circ$, and $90^\circ$, suggesting that the HMI rotated-image tests did not sample a full-enough rotation space to provide a clear test of DopHalf performance.

Unfortunately, we do not currently have {\em time-series} data of mean-value DopHalf results for input Dopplergrams rotated through a set of image orientations.

     \subsection[\textcolor{blue}{RingPhase Performance}]{\textcolor{blue}{RingPhase Performance}}
     \label{Results_RingPhase}

The P-angle analysis of the solar Dopplergrams using the RingPhase algorithm is mathematically more intensive than using the DopHalf algorithm, and therefore requires somewhat more computational time to run.  Here we break the results of testing the RingPhase algorithm into two subsections.  In the first (\S\ref{Results_RingPhase_fulldisk}) we present results that are either taken from or averaged across the full solar-disk of data, providing a direct correlation to the DopHalf results that have been presented so far.  Then, in \S\ref{Results_RingPhase_versusR}, we break away from full-disk P-angle results to explore the ways in which the global radial structure of the Dopplergrams presents in P-angle 'orientation' and impacts the mean results, with a comparison also to DopHalf r-dependent results.

          \subsubsection[\textcolor{blue}{Full-disk Results}]{\textcolor{blue}{Full-disk Results}}
          \label{Results_RingPhase_fulldisk}

\begin{figure}
     \begin{center}
     \includegraphics[scale=0.19]{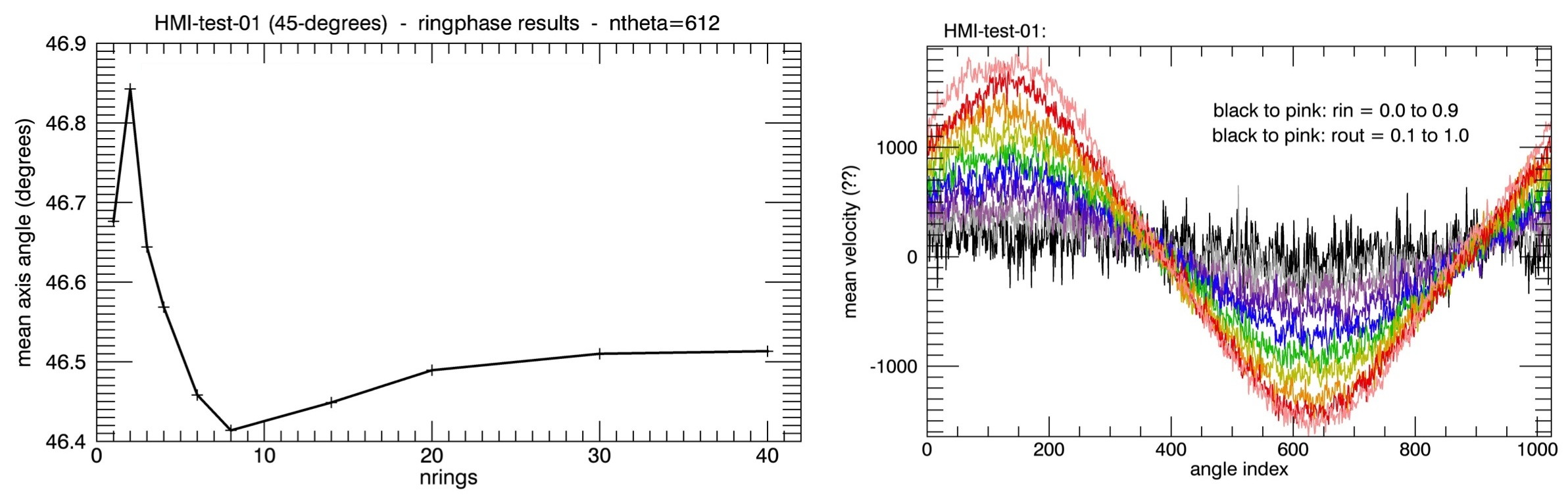}
     \caption[RingPhase full disk results as a function on $n_\mathrm{rings}$]{{\bf Left panel:} RingPhase results for HMI test image-1 (set rotation of $45^\circ$) averaged across the full disk as a function of the number of radial bins ($n_\mathrm{rings}$) used.  {\bf Right panel:} Sampled velocity curves for this test image similar to those used to compute the left-panel RingPhase results.}
     \label{FIG_Results_RingPhase_fulldiskVnrings}
     \end{center}
     \end{figure}

While we would like to report a full-disk value for the RingPhase computed P-angle, there is first the question of over how many annuli the Dopplergram-data should be sampled from which to compute a mean.  There is also the question of which solar radii contribute to the most stable computed signal, but this question is left for the next sub-section.  For now, the left-hand panel of Figure~\ref{FIG_Results_RingPhase_fulldiskVnrings}
presents the mean, full-disk RingPhase results from sampling a variable number of radial bins using one of the HMI test images.  This test image produces sampled velocity curves along the lines of those shown in the figure panel on the right.  We see here that the result of using two overlapping radial bins is about $0.35^\circ$ different than taking the mean of 40 (also overlapping) sample bins.  Using these results, we have selected $n_\mathrm{rings}=20$ as a reasonably well-converged run parameter for the rest of the RingPhase tests presented in this report.  Note that we have also chosen to use 612 angle bins for each annulus sample curve.  1024 and 2048 bins were also tested and were found to have a currently negligible deviation in results from the 612-bin computations.

Table~\ref{TAB_Results_RingPhase_HMIrot_values} presents the results of running RingPhase-full-disk-means on each of the HMI rotated test images.  The computed offsets from the known image rotations are similar to those computed by DopHalf (Table~\ref{TAB_Results_DopHalf_HMIrot_values}).  However, RingPhase does appear to be more stable to image orientation than DopHalf, in keeping with an algorithm that, in the theoretical limit, should be rotational-orientation independent.
\begin{table}
\caption{HMI-test Rotated-images RingPhase Results}
\centering
\begin{tabular}{c | c c }
	\hline \hline
	Rotation & 
	   RingPhase & $\delta$-RingPhase
	   \\ [0.5ex]
	\hline
	44	& 45.514	& 1.51	\\
	44.5	& 46.000	& 1.50	\\
	45	& 46.489	& 1.49	\\
	45.5	& 46.957	& 1.46	\\
	46	& 47.492	& 1.49	\\ [1ex]
	89.5	& 91.015	& 1.52	\\
	90	& 91.523	& 1.52	\\
	90.5	& 92.016	& 1.52	\\
	91	& 92.521	& 1.52	\\ [1ex]
	\hline
\end{tabular}
\label{TAB_Results_RingPhase_HMIrot_values}
\end{table}

\begin{figure}
     \begin{center}
     \includegraphics[width=\textwidth]{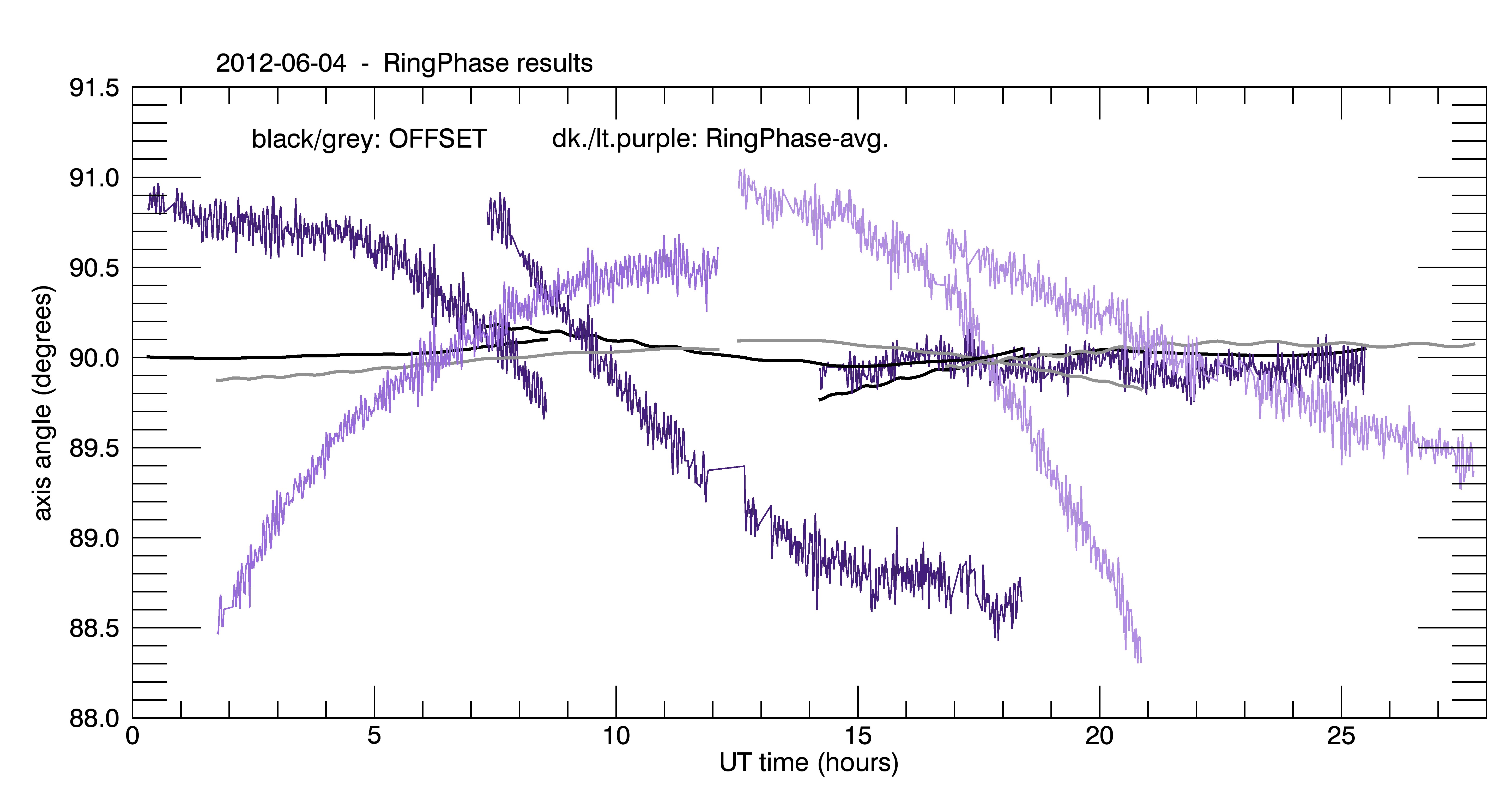}
     \caption[RingPhase results versus OFFSET across a full day]{Comparison of RingPhase results to input-file OFFSET values for all six sites on June $4^\mathrm{th}$, 2012.  Black OFFSET site-curves match to dark purple RingPhase curves, and grey OFFSET site-curves match to light purple RingPhase curves.}
     \label{FIG_Results_RingPhase_fulldisk_vnrings}
     \end{center}
     \end{figure}

Finally, Figure~\ref{FIG_Results_RingPhase_fulldisk_vnrings}
presents time-series results for full-disk-mean RingPhase run on a full-day's worth of observations for all six GONG sites.  Despite the improved image-orientation stability over DopHalf, the RingPhase curves also deviate by up to $1.5^\circ$ from the OFFSET best-P-angle estimate, with full-day results very similar to those produced by DopHalf (e.g., Figure~\ref{FIG_Results_DopHalf_Raw_vsOFFSET}).

          \subsubsection[\textcolor{blue}{Radial-structure Results}]{\textcolor{blue}{Radial-structure Results}}
          \label{Results_RingPhase_versusR}

\begin{figure}
     \begin{center}
     \includegraphics[width=\textwidth]{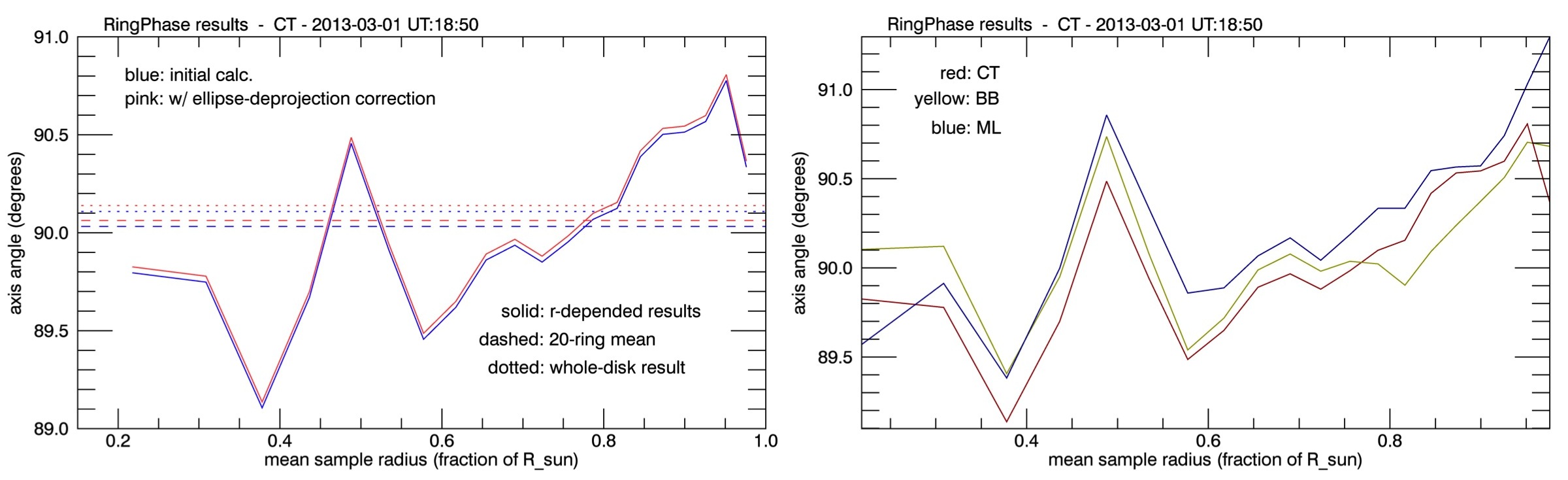}
     \caption[RingPhase results as a function of radius]{Example RingPhase results as a function of radius.  {\bf The left panel} compares one site's results to the ellipse-de-projection corrected results. 
     {\bf The right panel} compares concurrent results from three GONG sites.  Note that across these site observations, the suggested global structure is similar but not the same.}
     \label{FIG_Results_RingPhase_radial}
     \end{center}
     \end{figure}

Now we move on to explore the radial structure of P-angle-orientation results derived from solar Dopplergrams.  The left panel of Figure~\ref{FIG_Results_RingPhase_radial}
presents the r-dependent results of running RingPhase on a GONG-site Dopplergram collected at Cerro Tololo.  Both the ellipse-corrected and uncorrected curves are presented and, while the correction does produce a noticeable difference, we see that it is negligible compared to the nearly $2^\circ$ of P-angle range produced by the Dopplergram radial structure.  Furthermore, this global structure appears to produce somewhat of a gradient in P-angle results rather than variations about some fundamental value, so there does not appear to be a clear method, looking at the r-dependent results for a single observation, to pick out the 'best' value.

\begin{figure}
     \begin{center}
     \includegraphics[scale=0.22]{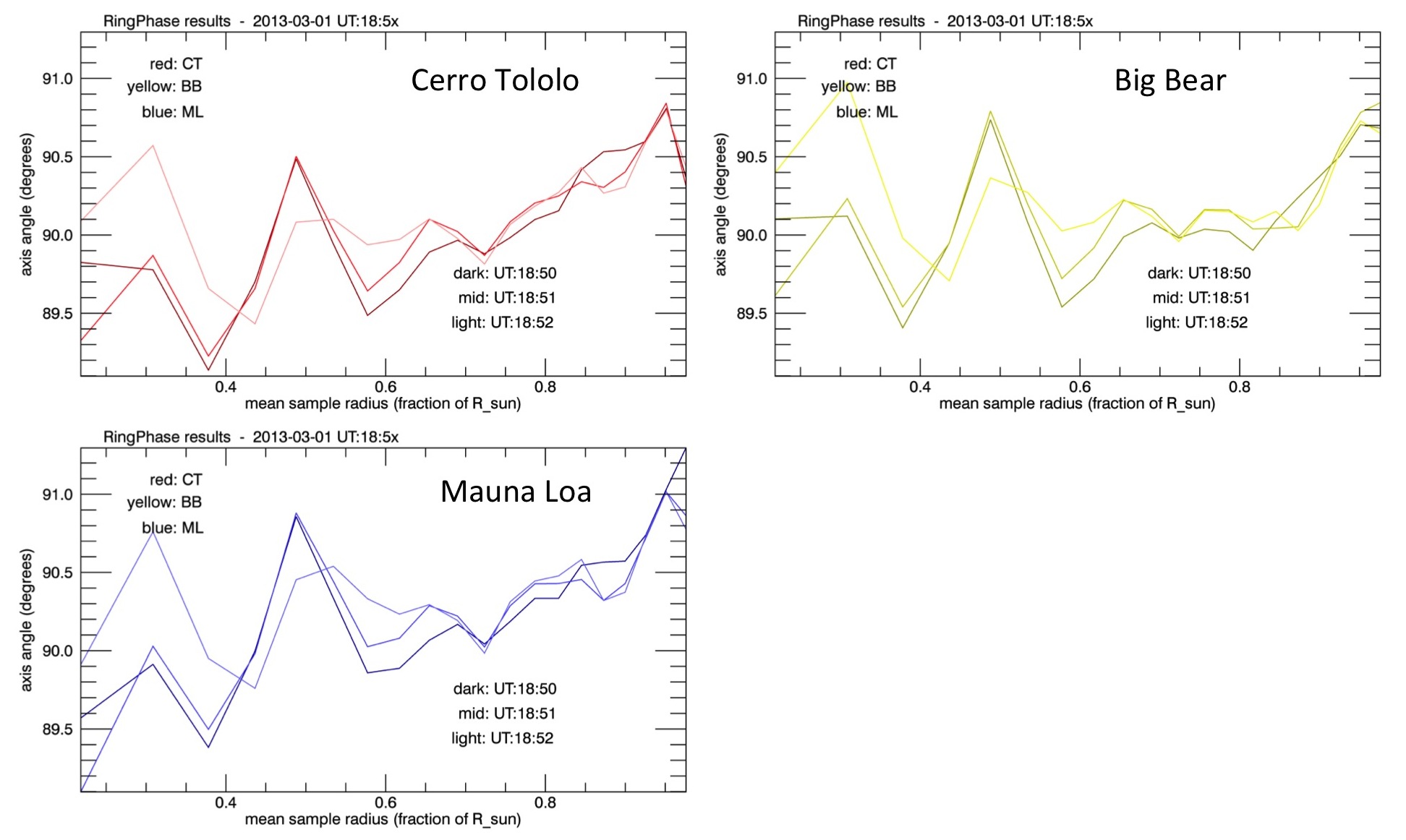}
     \caption[RingPhase r-dependent minute-to-minute results]{Example RingPhase r-dependent minute-to-minute results for three sites.  Note that certain radii appear to be more temporally stable than others, e.g., near $r=0.73$.}
     \label{FIG_Results_RingPhase_radial_VtimeVsites}
     \end{center}
     \end{figure}

\begin{figure}
     \begin{center}
     \includegraphics[width=\textwidth]{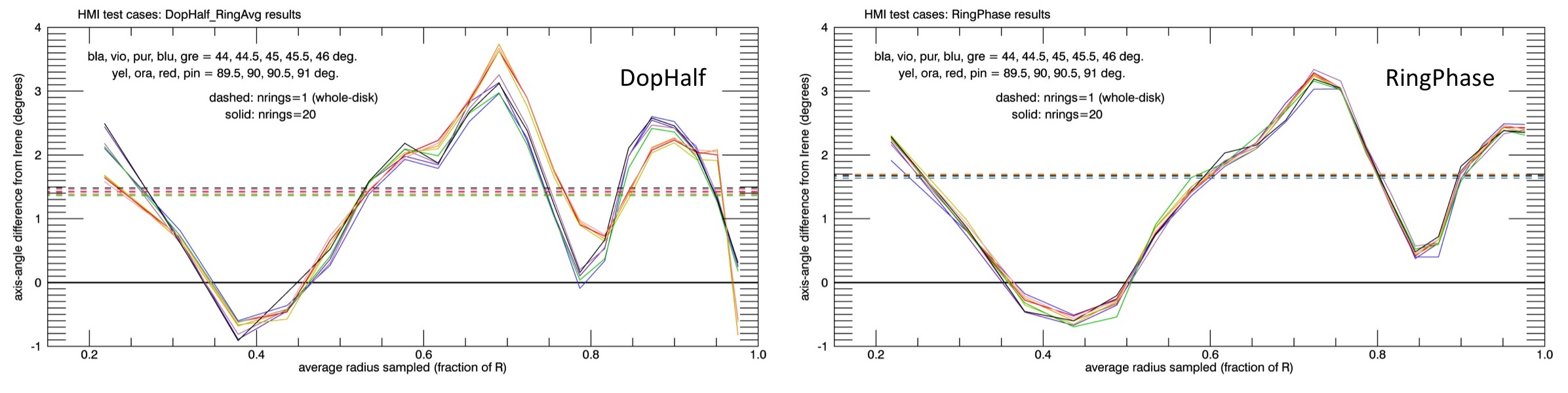}
     \caption[R-dependent HMI test results for DopHalf versus RingPhase]{HMI-test results (for an image at several rotated orientations) for DopHalf versus RingPhase as a function of radial-annulus bin.  The results plot the computed P-angle minus the imposed rotation.  Note that RingPhase appears to be somewhat more stable to image orientation, but still shows a distinct spread.}
     \label{FIG_Results_RingPhase_radial_DopHalfVRingPhase}
     \end{center}
     \end{figure}

However, in the {\em right} panel of Figure~\ref{FIG_Results_RingPhase_radial} we do see that at least some of the radial structure may be detectibly consistent from site-to-site, most noticeably in the mid-range of solar radius.  Or, in Figure~\ref{FIG_Results_RingPhase_radial_VtimeVsites}
we see that some locations may have more short-term temporal stability than others, though from this example set these appear to be more likely to occur in the outer-half of solar radii.  Finally, looking again at the HMI set of test data, we do see in Figure~\ref{FIG_Results_RingPhase_radial_DopHalfVRingPhase}
that DopHalf and RingPhase produce very similar radial-structure results.  However:
\begin{enumerate}
     \item As before, DopHalf shows greater variation between rotated images results.
     \item Curiously, the DopHalf-structure results appear to be radially shifted inward relative to the RingPhase results.
\end{enumerate}

\begin{figure}
     \begin{center}
     \includegraphics[scale=0.265]{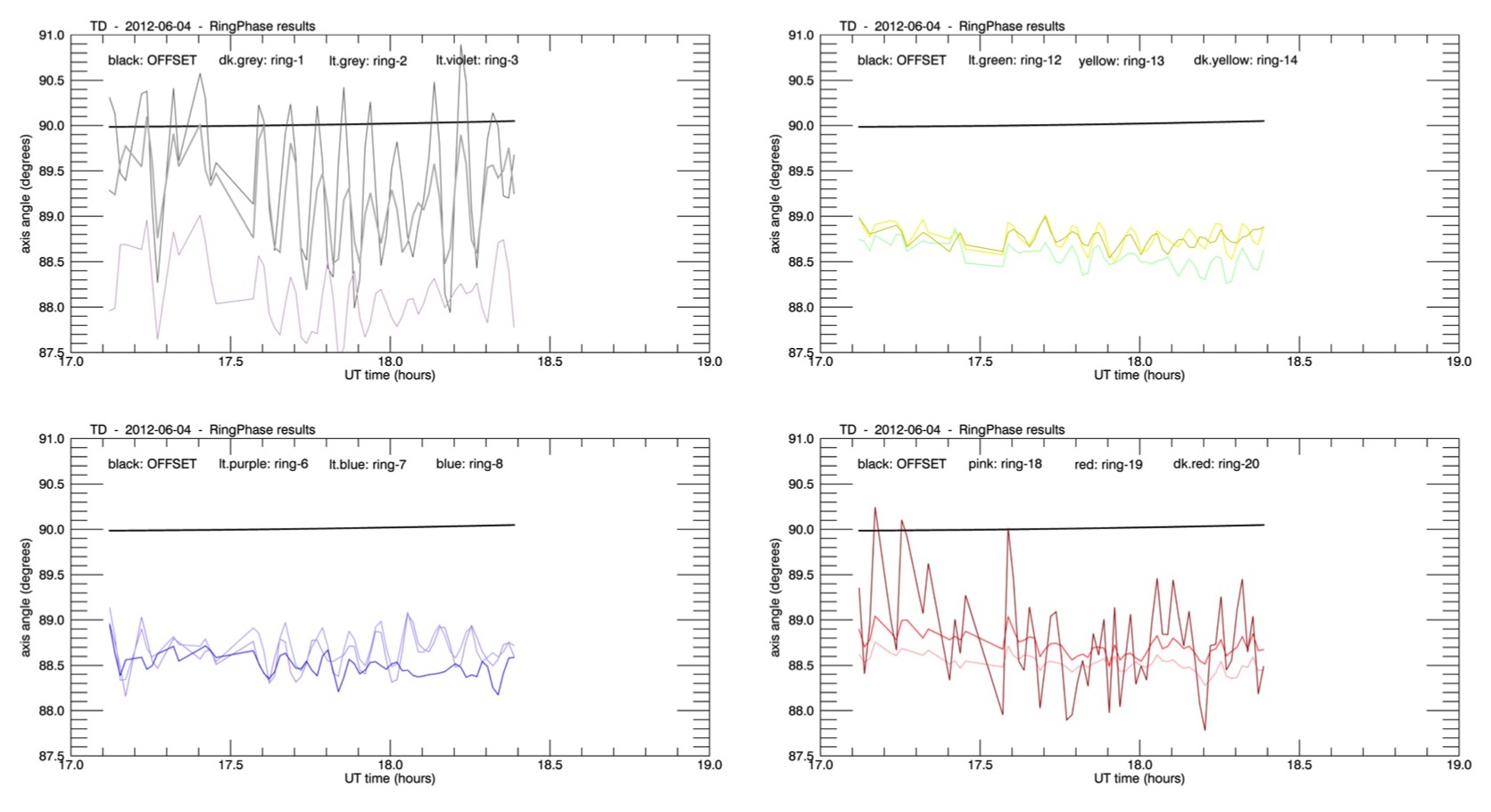}
     \caption[Minute-to-minute RingPhase results for different radial annuli bins]{Example minute-to-minute RingPhase results for a set of TD observations for different radial bins.  A subset of annulus rings are presented here.  They are equal-area and the full set is numbered from 1 (innermost disk) to 20 (outmost limb).  Note that the results appear to be more stable in the outer-half of the disk (excluding the outermost ring of the limb).}
     \label{FIG_Results_RingPhase_radial_Vtime}
     \end{center}
     \end{figure}

Taking a closer look then at the temporal stability in the results produced at different solar radii, we present Figure~\ref{FIG_Results_RingPhase_radial_Vtime}.
Here we plot a time-series of RingPhase results along separate curves for each sampled annulus bin.  These plots clearly suggest that the innermost and outermost bins produce unstable results, and that better stability is likely to be found if one excludes approximately the inner half of the solar-disk Dopplergram.

     \subsection[\textcolor{blue}{Versus OFFSET and Velocity Gradients}]{\textcolor{blue}{Versus OFFSET and Velocity Gradients}}
     \label{Results_VelGrad}

\begin{figure}
     \begin{center}
     \includegraphics[scale=0.24]{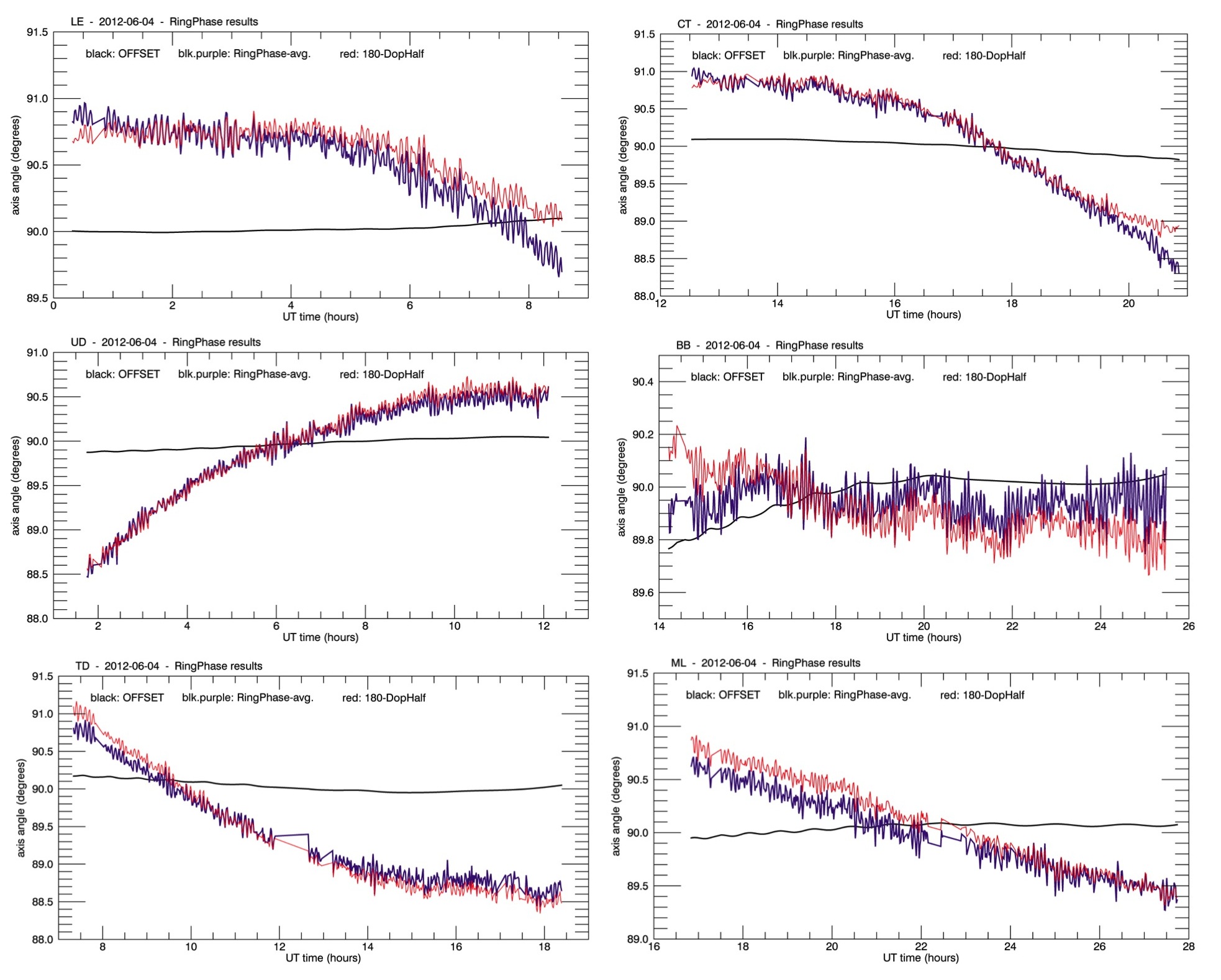}
     \caption[Comparison of RingPhase and DopHalf]{Comparison of RingPhase and DopHalf versus OFFSET for all six sites for June 4,  2012 .}
     \label{FIG_Results_RingPhase_WDopHalfVOFFSET}
     \end{center}
     \end{figure}

In this section, we take a more general look at the use of GONG-site Dopplergrams for estimating P-angle.  As can be seen in Figure~\ref{FIG_Results_RingPhase_WDopHalfVOFFSET},
both the RingPhase and DopHalf algorithm produce very similar results, especially in terms of full-day curves for each site.  At some fundamental level, this is the P-angle-type information that is available within the Dopplergram data.  However, it does not match well to the OFFSET-reported P-angle value.

\begin{figure}
     \begin{center}
     \includegraphics[width=\textwidth]{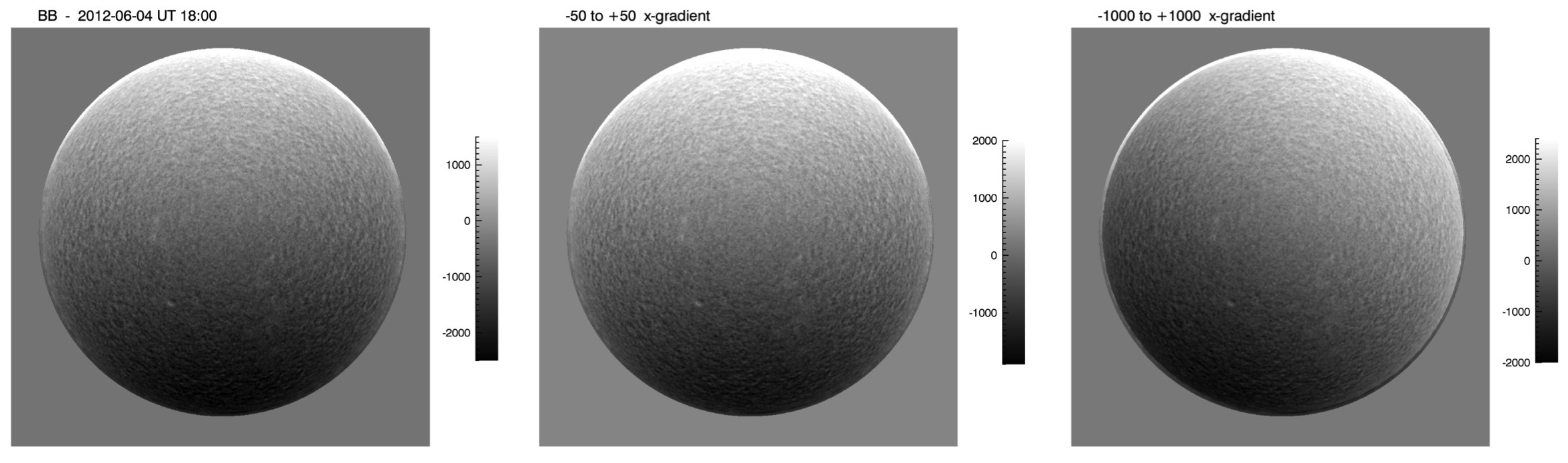}
     \caption[Dopplergrams with added velocity-x-gradients]{A Dopplergram taken at Big Bear both as is and with the addition of two example \\ x-direction gradients.}
     \label{FIG_vgradDopplergrams}
     \end{center}
     \end{figure}

There are likely a large number of reasons why this might be.  Here we would like to take a look at one (probably over-simplified) possibility: site-to-site variations in the Dopplergram results, specifically, the potential for an added gradient in reported Doppler velocity across the image.  In Figure~\ref{FIG_vgradDopplergrams},
we present the visual results of adding an x-direction gradient to an example Dopplergram image.  By visual inspection, a gradient going from -50 at one limb to +50 at the other is not particularly noticeable, except perhaps in the overall scaling of the data.  However, a gradient that goes from -1000 to +1000 produces a Dopplergram image that appears to have a P-angle shifted nearly $45^\circ$.

In Table~\ref{TAB_Results_VelGrad}
\begin{table}
\caption{del Teide 2012-06-04 UT 18:00 Dopplergram results with added x-gradient}
\centering
\begin{tabular}{c | c | c }
	\hline \hline
	$-$ to + gradient peaks & 
	   DopHalf & 
	   RingPhase
	   \\ [0.5ex]
	\hline
	1000		& 116.04	& 116.05	\\
	100		& 91.5	& 91.43	\\
	10		& 88.86	& 88.77	\\
	1		& 88.59	& 88.51	\\ [1ex]
	0		& 88.56	& 88.48	\\ [1ex]
	-1		& 88.53	& 88.45	\\
	-10		& 88.27	& 88.18	\\
	-100		& 85.63	& 85.54	\\
	-1000	& 61.68	& 61.56	\\ [1ex]
	\hline
	OFFSET = 90.02$^\circ$ \\ [1ex]
	\hline
	50		& 		& 90.04$^\circ$	\\
	53		& 90.04$^\circ$	& 		\\ [1ex]
	\hline
\end{tabular}
\label{TAB_Results_VelGrad}
\end{table}
we present the results of running DopHalf and RingPhase on an example Dopplergram that has first been modified using a range of x-direction gradients.  In this example, the original-Dopplergram results report a P-angle of about $88.5^\circ$ versus an OFFSET angle of $90.02^\circ$.  By adding a gradient across the Dopplergram of around $\pm 50$ m/s, the Dopplergram results become $90.04^\circ$.

\begin{figure}
     \begin{center}
     \includegraphics[width=\textwidth]{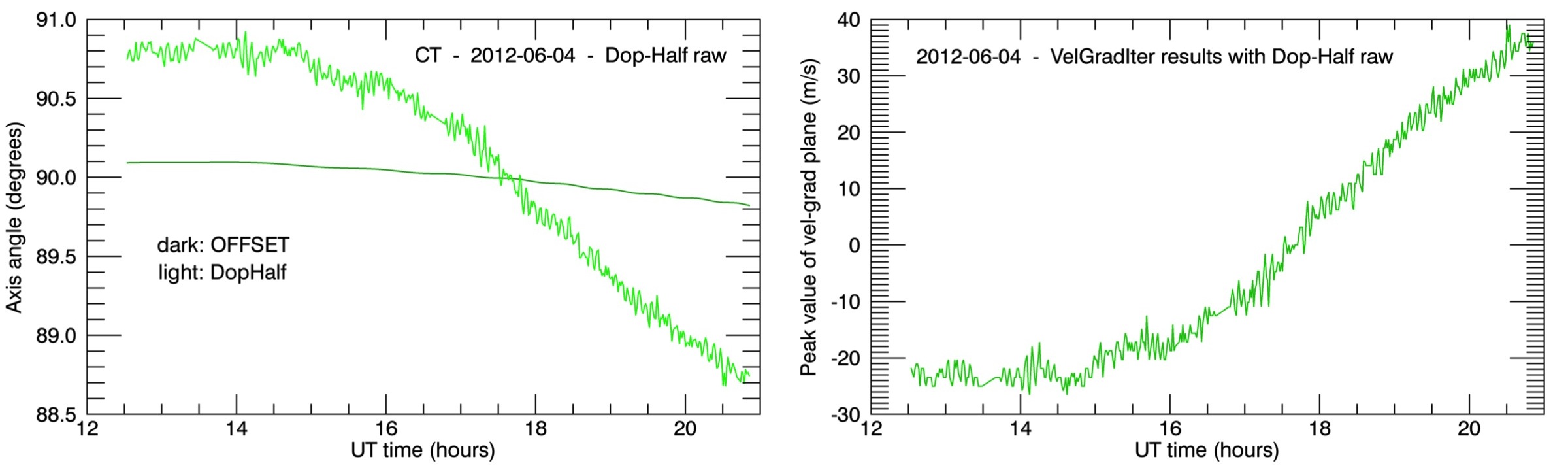}
     \caption[Comparison of DopHalf results to x-Velocity-gradient 'corrections']{The {\bf left panel} presents the OFFSET values and DopHalf results (using raw Dopplergrams) for a full day of Cerro Tololo data.  The {\bf right panel} presents the results of iteratively searching, for each observation, for an x-direction gradient value to add across the solar Dopplergram such that newly computed DopHalf results match the reported OFFSET values to within $0.02^\circ$.}
     \label{FIG_VelGradIter_fullday}
     \end{center}
     \end{figure}

However, extrapolating this OFFSET-matching-via-x-gradient idea to a time-series reveals the flaws of this sort of analysis.  In Figure~\ref{FIG_VelGradIter_fullday}
we present the OFFSET curve and raw-Dopplergram DopHalf results for a full day of Cerro Tololo data in one panel.  In the second panel are the results of iteratively sampling different x-direction gradient values overlain onto each observation until the computed DopHalf value matches the reported OFFSET value.  From this set of curves, it is clear that the matching-gradient value is very much a function of the original DopHalf value, without any clear underlying preference for some gradient value that might point to instrumental effects.  This does not mean that instrumental effects are not at issue, only that a quick-fix sort of analysis is insufficient to characterize them.

\section[\textcolor{blue}{Current Conclusions}]{\textcolor{blue}{Current Conclusions}}
\label{Conclusions}

The conclusions reached by the current report are these:
\begin{enumerate}
     \item That despite their very different analytical approaches, both of the Dopplergram-based P-angle-estimating algorithms, DopHalf and RingPhase, fundamentally agree on the shape and location of the P-angle curve throughout an observing day for a particular site.
     \item That there is (and quite possibly always will be) significant temporal noise in estimating the P-angle via a site Dopplergram due to things like p-mode oscillations.
     \item That, noise levels aside, the Dopplergram methods do {\em not} do a good job of reproducing the OFFSET curves.  However:
     \item The variation between OFFSET and the Dopplergram curves appears to be strongly site/instrument dependent.  If the variation were fundamentally due to the Dopplergram expression of physical changes in the solar surface, then there should be some sort of concurrent matching of curve offsets from one site to another.  There are no such clear correlations.  Therefore:
     \item Further work needs to be done to better understand the processing of and influences on the individual GONG-site observations before a clear evaluation of these Dopplergram methods can be made.  Until it is better understood the degree to which these site-to-site differences are intrinsic to the observations/processing versus, for example, seeing dependent, these algorithms need to be set aside.
     \item If the site-to-site data differences are eventually found to be primarily intrinsic and correctable, the minute-to-minute correlations in these P-angle results between sites leave open the possibility that these algorithms may yet prove useful.
\end{enumerate}

\section[\textcolor{blue}{Future Work}]{\textcolor{blue}{Future Work}}
\label{Future}

As with all research projects, there are a large number of directions in which future headway might be made.  Therefore, this section on potential future work is broken up into two subsections.

In \S\ref{Future_Tests} we discuss direct extensions to the algorithms and tests presented in this report that we would like to followup on.  
And in \S\ref{Future_Algorithms} we discuss possibilities for alternative algorithms that might be worth exploring at a later date.

Before those discussions, however, is the reminder that there is one important question that needs to be addressed before further progress can be made on DopHalf- and RingPhase-type algorithms, which is: what is the level of calibration of the site Dopplergram values?  Unfortunately, this is a tricky question to answer.  However, while Figure~\ref{FIG_Results_RingPhase_WDopHalfVOFFSET} makes it clear that there {\em are} site-dependent variations between the Dopplergram-P-angle-type curves and the reported OFFSET curves, from Figure~\ref{FIG_Results_DopHalf_Filtered_toppDopHalfs_days} it is also clear that there {\em is} consistency from day-to-day in the measured DopHalf curves at a given site.

The results of \S\ref{Results_VelGrad} show that a simple x-gradient is insufficient to describe whatever may be at issue with the site Dopplergrams.  But, of course, because the GONG data are recorded in full x-by-y images, that result is not too surprising.  Also, those gradients were applied {\em after} the VELSCALE and VEL\_BIAS corrections, so there was already a certain degree of separation from the raw data.

In a separate study by Clark, et al.~\cite{ClarkHarveyHillToner2003}, the authors describe a zero-point correction to the GONG magnetograms that they identify via inspection and fitting of quiet regions across the solar image.  However, they also encountered a time-varying component to the zero-point offset linked to the operation of the instrument.  Unfortunately, Dopplergrams do not have 'quiet regions' where the velocity is expected to measure near zero.  Therefore, the best way forward on the issue of calibration remains unclear.  Perhaps there is some clue in the right-hand panel of Figure~\ref{FIG_Results_RingPhase_radial}, which depicts some similarities, but also several points of variation, between the RingPhase-sampled radial structure of concurrent Dopplergrams taken from three separate sites.

     \subsection[\textcolor{blue}{Further DopHalf and RingPhase Extensions and Tests}]{\textcolor{blue}{Further DopHalf and RingPhase Extensions and Tests}}
     \label{Future_Tests}

According to the tests performed so far, there are a couple of areas where the DopHalf and RingPhase algorithms could be adjusted for likely or possible improvement.  The first is in the solar sampling radius.  The results presented in Figure~\ref{FIG_Results_RingPhase_radial_Vtime} suggest that excluding the inner half in area of the solar Dopplergram along with the outermost rim may lead to noticeably less noisy P-angle results.  This should likely hold true for RingPhase and for DopHalf using raw Dopplergrams, but would be less likely to reduce noise in the DopHalf results using filtered Dopplergrams, since the entire image is used to generate the filter.

Second, while the RingPhase results should be reasonably stable to image orientation, the DopHalf results demonstrably are not.  It seems likely that incorporating some sort of a rotated-image average into the DopHalf algorithm would provide more accurate results.  However, there is still a fair bit of testing that would need to be done in order to determine what sort of rotated average would be best.  Further, such testing would/will be much more effective once the calibration status of the GONG-site Dopplergrams is better understood.  Such testing should involve the inspection of time-series results and should probably take a look at:
\begin{enumerate}
     \item The variability of the amplitude of the rotation curve for each Dopplergram (i.e., the max-min of curves like those shown in Figure~\ref{FIG_Results_DopHalf_rot_4sitescurves}).
     \item The mean results produced by rotated raw Dopplergrams.
     \item The mean results produced by rotated filtered Dopplergrams.
     \item The results produced by FFTwide-filtered Dopplergrams at the orientation where the two FFT-component phase angles are closest to $\pm 90^\circ$.
\end{enumerate}

Also, for DopHalf there is the question of if/how-much the B0-angle of the observed sun affects the results (for RingPhase B0 almost {\em certainly} affects the results).  Probably the simplest way to test this question would be to produce a series of synthetic, smooth images using a model Dopplergram of a  perfectly rotating body.  The image series should include both a range in P-angle orientation as well as a range in modeled B0-angle, in order to get a sense of to what extent, and possibly under what P-angle dependence, B0 alters the DopHalf results.  Unfortunately, even without surface structure, the sun is not a perfect rotator, so the results of this test would be limited by the degree to which the model rotator matches the solar rotation profile.

Finally, there are tantalizing anecdotes to suggest that aligning the fine-structure of the time-varying results curves produced by either DopHalf or RingPhase, or possibly matching up stable-radial-structure points between GONG sites, could {\em maybe} produce nice, tight image-alignment-worthy difference angles.  This is probably a pipe-dream, but if everything else is looking fabulous, it might be worth looking into again.

     \subsection[\textcolor{blue}{Other Potential Algorithms}]{\textcolor{blue}{Other Potential Algorithms}}
     \label{Future_Algorithms}

The development of the algorithms presented in this report has lead to ideas for {\em other} algorithms that might be potentially worth pursuing.  The first is an algorithm to potentially mitigate the effect of the solar B0-angle in especially the RingPhase but possibly also the DopHalf results.  The second is an idea for a new Dopplergram-based P-angle algorithm that should not be vulnerable to variations in observed B0.  Both of these algorithms require iterative solutions and so would be more computationally intensive than the algorithms tested so far.

The {\bf B0-correction algorithm} would be an iterative-adjustment algorithm incorporated into the functionality of the DopHalf and/or RingPhase codes.  This algorithm would:
\begin{enumerate}
     \item Begin with an initial guess of the P-angle (either supplied by DopHalf or using the assumption of $\theta_P \sim 90^\circ$).
     \item Using this axis angle and the known B0-angle of the sun, redefine the sample-area of the solar disk as shown in Figure~\ref{FIG_Future_B0correct},
     \begin{figure}
     \begin{center}
     \includegraphics[scale=0.18]{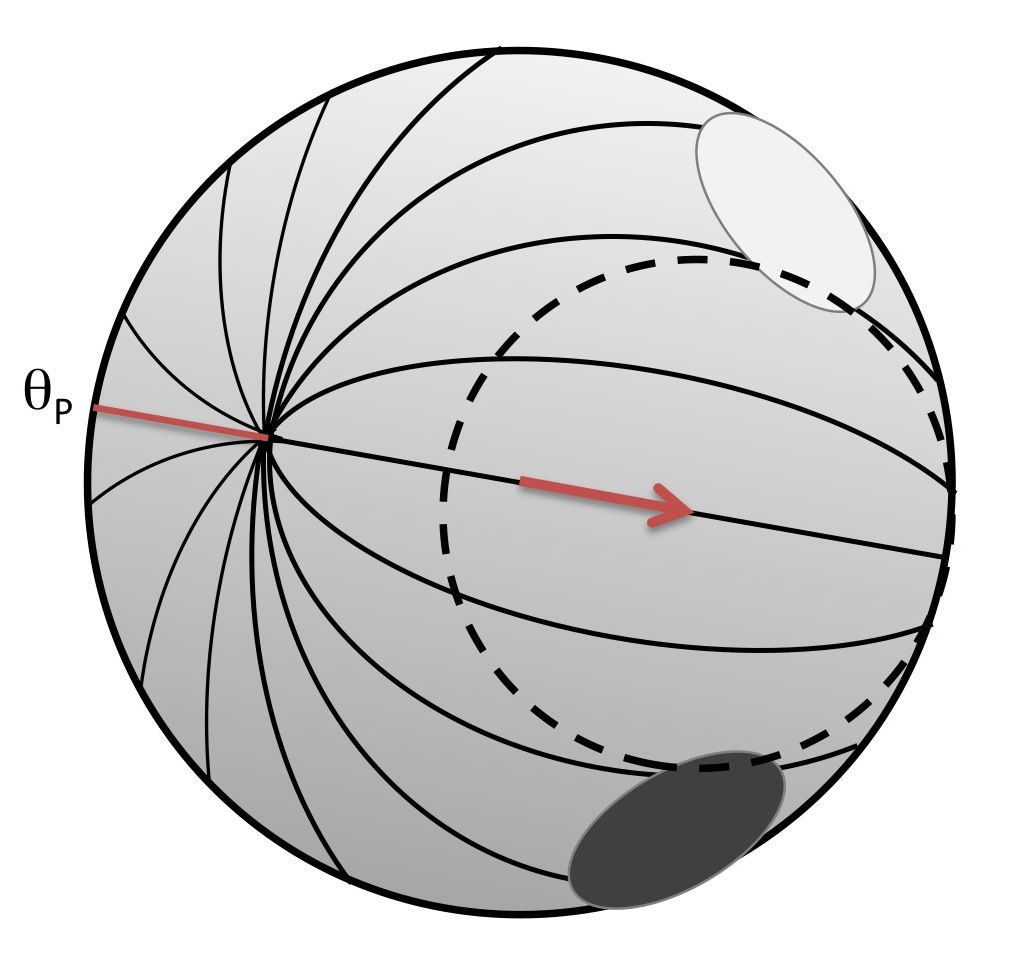}
     \caption[Cartoon of B0 correction]{A cartoon (for an observation with an exaggerated B0) of re-defining the disk sample area to mitigate the effect of the observed B0-angle.}
     \label{FIG_Future_B0correct}
     \end{center}
     \end{figure}
     such that the radius of the new sample disk is reduced by $R_\odot \left[ 1 - \sin\left(B0\right) \right]$, and the disk-center location is shifted downward along the P-angle-defined axis by the same amount.
     \item Using the newly defined sample disk, calculate the P-angle using the primary algorithm of interest.
     \item Use the new P-angle to shift the center of the sub-disk sample area, re-calculate the P-angle, and continue iteratively until the P-angle is defined to within sufficient tolerance.
\end{enumerate}
As this algorithm has not yet been explored, it is not clear how well it would or would not correct for B0 errors.  However, at the very least it should allow RingPhase to examine sample areas with min/max velocity peaks much closer to $\pm 90^\circ$ from the pole positions than otherwise.

The {\bf Statistical-symmetry algorithm} is an idea for an entirely different Dopplergram-based P-angle algorithm that would be independent of the rotational and viewing geometries of the sun.  The implementation of this algorithm would be purely a search for an axis of symmetry in the Dopplergram data.  To accomplish this, the algorithm would:
\begin{enumerate}
     \item Adjust the Dopplergram values in the image by calculating the mean velocity value on the disk and adding a global offset to the data so that this mean value is 0.0.  Then take the absolute values of the computed velocities.
     \item Begin with an initial guess of the axis angle, as with the B0-correction algorithm discussed above.
     \item Use the axis angle and reported sun-center location to define an axis running across the solar disk.  From this axis, compute value pairs for points at equal-latitude-along and equal-distance-from this axis.  This would likely require computing an interpolated value on one side of the image to match against a pixel value on the other.
     \item Using these value pairs, compute a statistical measure of the symmetry displayed along this axis, possibly with a sum in the difference between values across the image.
     \item Varying the axis angle from the initial guess, recompute the symmetry measure, and continue iteratively until an axis angle is found that maximizes symmetry (perhaps by minimizing the sum of difference values between pairs).
\end{enumerate}
This algorithm would clearly be vulnerable to any errors in the reported sun-center positions, but that is actually true of any P-angle-finding algorithm.  It would also suffer if the initially computed Doppler-velocity mean were noticeably off from the appropriate rest-velocity for the observation, such that the computed difference-values between symmetry points are forced to be shifted away from zero even for the correct P-angle axis.  Both of these errors would likely sample a correct axis with  a small {\em spread} in difference values, but with a mean difference $\ne$ 0.0.

$\,$

\renewcommand{\abstractname}{Acknowledgements}

Authors acknowledge the pioneering work carried out by Cliff Toner which was instrumental in setting up a scheme for P-angle alignment in GONG Dopplergrams.   We would also like to thank Jack Harvey,  Frank Hill and Philip H. Scherrer for helpful comments. This work utilises GONG data obtained by the NSO Integrated Synoptic Program, managed by the National Solar Observatory, which is operated by the Association of Universities for Research in Astronomy (AURA), Inc. under a cooperative agreement with the National Science Foundation and with contribution from the National Oceanic and Atmospheric Administration. The GONG network of instruments is hosted by the Big Bear Solar Observatory, High Altitude Observatory, Learmonth Solar Observatory, Udaipur Solar Observatory, Instituto de Astrof\'{\i}sica de Canarias, and Cerro Tololo Interamerican Observatory.

\end{document}